\documentclass[reprint,showpacs,aps,pra,amsmath,amssymb]{revtex4-1}

\usepackage{graphicx,epstopdf}
\usepackage{dcolumn}
\usepackage{bm}
\usepackage{amsmath,empheq}
\usepackage{color}
\usepackage[bookmarks]{hyperref}
\usepackage{natbib}
\usepackage{CJK}

\allowdisplaybreaks[2]

\begin{document}
\begin{CJK*}{UTF8}{gbsn}

\title{Nonlocal nonlinear response of thermal Rydberg atoms and modulational instability in absorptive nonlinear media}

\author{Lida Zhang(\CJKfamily{gbsn}张理达)}

\author{J\"{o}rg Evers}
\affiliation{Max-Planck-Institut f\"{u}r Kernphysik, Saupfercheckweg 1, D-69117 Heidelberg, Germany}

\date{\today}

\begin{abstract}
Nonlinear and nonlocal effects are discussed in the interaction of laser fields with thermal Rydberg atoms in electromagnetically induced transparency configuration. We show that under the crucial approximation that the time variation in the dipole-dipole interactions due to atomic motions can be neglected in an ensemble average, an analytical form can be obtained for the nonlocal nonlinear atomic response of the thermal medium, and study it for different parameter cases. 
We further propose a generalized model to describe the modulational instability (MI) in absorptive nonlinear media, in order to understand the propagation dynamics in the thermal Rydberg medium. Interestingly, this model predicts that at short propagation distances, each wave component exhibits the MI effect in absorptive nonlinear media, unlike in the purely dispersive case.

\end{abstract}

\pacs{32.80.Ee, 42.50.Gy, 42.65.-k, 34.20.Cf}

\maketitle

\end{CJK*}

\section{Introduction}

Rydberg atoms, possessing large dipole moments and therefore long-range dipole-dipole interactions (DDI), have attracted intensive interests due to its potential applications in diverse fields~\cite{Saffman2010RMP,review-1,review-2}. The last decade has witnessed great progresses in the study of ultracold Rydberg atoms. On the one hand, a series of beautiful experiments have shown that ultracold Rydberg atoms are capable to realize photon blockade~\cite{Gorshkov2011PRL,Dudin2012Science,Baur2014PRL}, interacting photons~\cite{Peyronel2012Nature,Firstenberg2013Nature}, single-photon transistor~\cite{Gorniaczyk2014PRL,Gorniaczyk2015arXiv}, transport phenomena~\cite{Schempp2015PRL,Guenter2013Science}, to mention a few. On the other hand, several theoretical models have also been proposed to successfully describe the interacting ultracold Rydberg atoms under certain circumstances, including mean field models~\cite{Tong2004PRL,Weimer2008PRL}, cluster expansions~\cite{Schempp2010PRL,Heeg2012PRA}, rate equations~\cite{Ates2007PRA,Vincent2013PRL,
Gaerttner2014PRA}. In those models, the thermal motion effects is reasonably neglected at typically short time scale~($\sim\mu s$) and low temperature~($\sim\mu \text{K}$) due to the dominant role of many-body coherent dynamics in dilute ultracold Rydberg atoms.

In parallel to these developments in ultracold Rydberg atoms, first experiments have been conducted in thermal Rydberg gases~\cite{Raimond1981JPB,Shen2007PRL,Kubler2010NPhoto,Huber2011PRL,kolle2012PRA,Sedlacek2012NPhys,Baluktsian2013PRL,Li2013PRL,Brredo2013PRL,Carr2013PRL,Urvoy2015PRL}, demonstrating evidence for long-range coherent DDI. In these experiments, typically two off-resonant probe and control fields interact with the thermal atoms in such a way that the two-photon transition to the Rydberg state is near-resonantly driven, while the single-photon transitions are far off-resonance. In this electromagnetically-induced-transparency (EIT) configuration, the theoretical description can be simplified~\cite{Sevincli2011PRL} since the time evolution of the intermediate state can be adiabatically eliminated. However, this simplified picture cannot fully describe the involved dynamics fo the thermal Rydberg ensemble. The greatest challenge is posed by the time-varying DDI in the thermal regime due to the 
atom motion, but also other effects such as 
Doppler shifts and atomic collisions could lead to a breakdown of present theoretical models.

\begin{figure}[b]
\includegraphics[width=3.0cm]{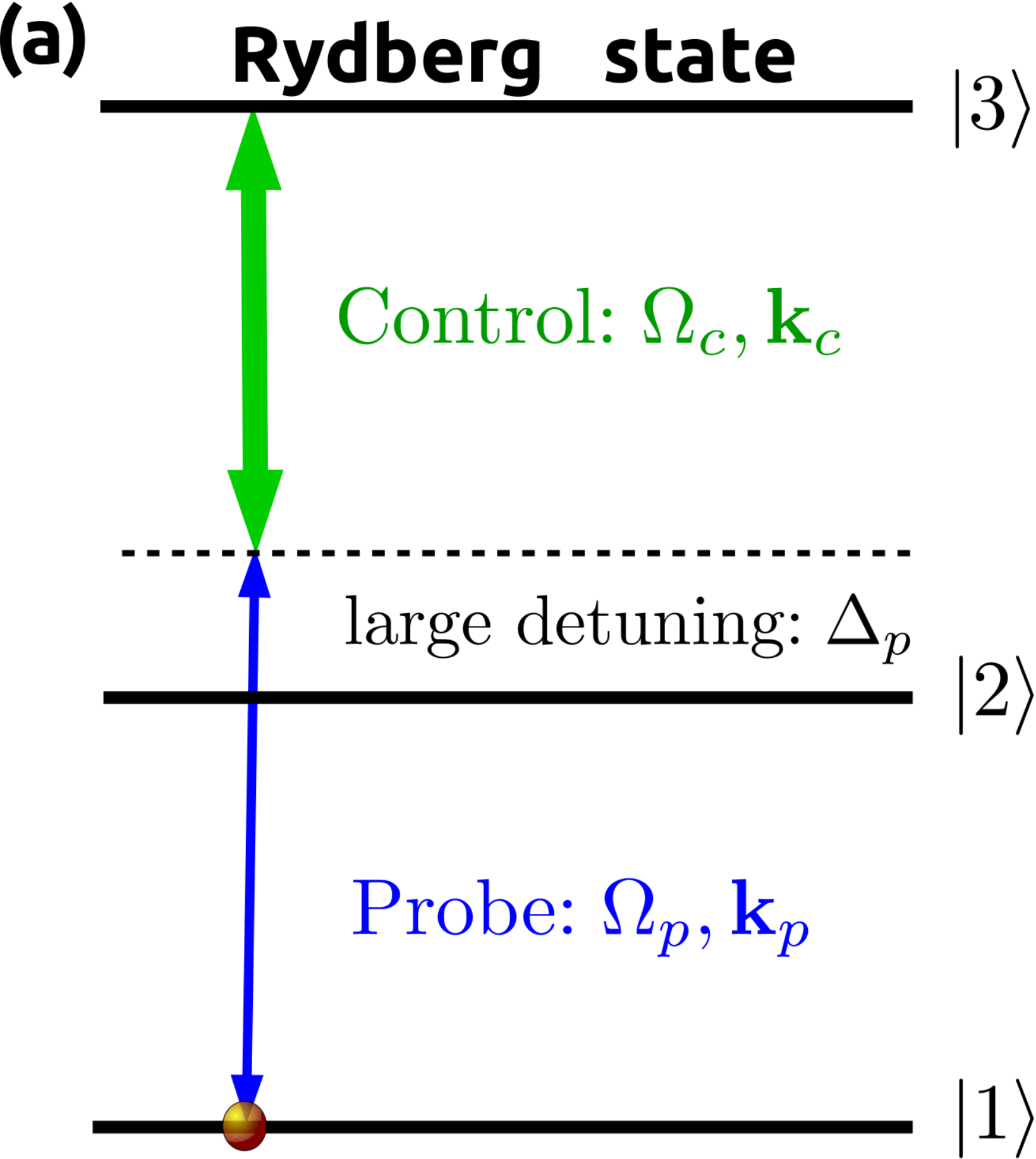}\hspace{3em}%
\includegraphics[width=4.5cm]{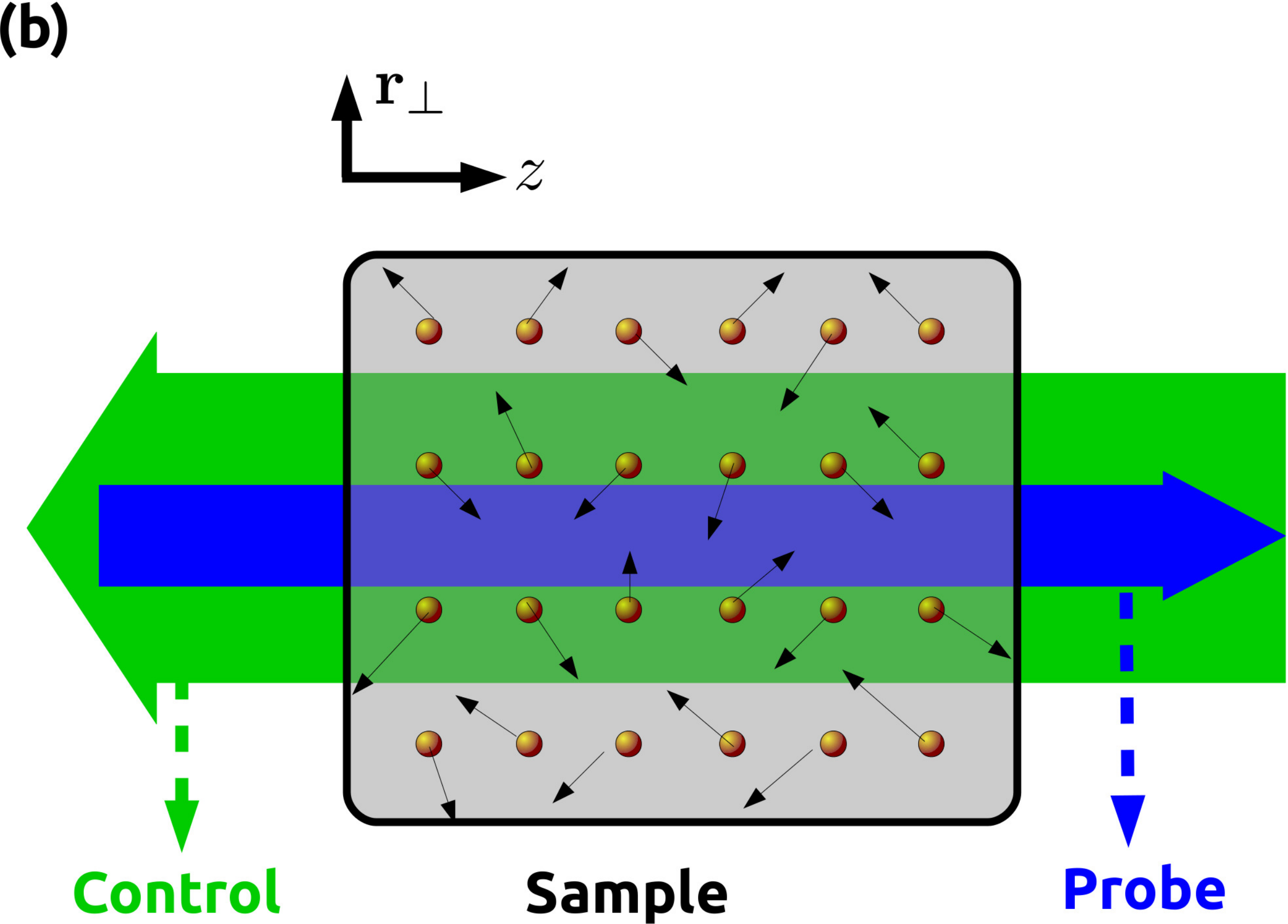}
\caption{(Color online) The three-level scheme for a single atom (a) interacting with counter-propagating probe and control fields (b). The two single-photon transitions $|1\rangle\leftrightarrow|2\rangle$ and $|2\rangle\leftrightarrow|3\rangle$ are driven by the two far-detuned probe and control fields with Rabi frequencies $\Omega_{c}$ and $\Omega_{p}$ respectively, while the two-photon transition from $|1\rangle$ to the high-lying Rydberg state $|3\rangle$ is of near resonance.}
\label{fig1}
\end{figure}

Here, we develop a different approach to interacting thermal  Rydberg atoms. This model still essentially neglects the temporal variations of the DDI due to the atom motion, but apart from that provides a proper description of the steady state of thermal Rydberg atoms interacting with two laser fields in EIT configuration. In particular, we derive analytical expression for the nonlinear nonlocal response of the thermal Rydberg atoms. Based on this result, we find that in the single-photon near-resonant regime the nonlinear absorption for the probe field is weakened as the temperature increases. On the contrary, when the laser fields are far-off resonance with the single-photon transitions, the nonlocal nonlinear dispersion remains almost unchanged while the nonlocal nonlinear absorption is firstly enhanced and then weakened as the temperature grows. 
Motivated by these results,  we further introduce a generalized model for the modulational instability (MI) in absorptive nonlinear media. Analytical solutions from this model agree well with numerical simulations of the propagation dynamics. In contrast to purely dispersive nonlinear media, this model predicts that at short propagation distance, each wave component exhibits the MI effect. At longer propagation distances, however, for most frequency components MI competes with both, the nonlocal nonlinear absorption and coupling to other spatial frequency modes, and eventually is suppressed.

The paper is organized as follows. In Sec.~\ref{sec2}, we introduce our model for the interacting thermal Rydberg atoms and derive the nonlinear atom response. In Sec.~\ref{dynamics}, we numerically solve the paraxial wave equation to study the probe field  propagation dynamics, and  present our main results. Section~\ref{summary} contains a discussion and summary of our results.

\section{\label{sec2}Theoretical considerations}

\subsection{\label{sec:model} Theoretical model}

We consider a ladder-type three-level atomic system as shown in Fig.~\ref{fig1}. The ground state $|1\rangle$ is coupled by a weak probe field $\Omega_{p}$ with frequency $\omega_{p}$ and wavevector $\textbf{k}_{p}$ to the intermediate state $|2\rangle$, which is further driven by a strong control field $\Omega_{c}$ with frequency $\omega_{c}$ and wavevector $\textbf{k}_{c}$ to a high-lying Rydberg state $|3\rangle$. The two field frequencies are chosen such that the two-photon transition from $|1\rangle$ to $|3\rangle$ is driven at near-resonance, thus forming an electromagnetically induced transparency (EIT) configuration. In contrast, the single-photon transitions $|1\rangle\leftrightarrow|2\rangle$ and $|2\rangle\leftrightarrow|3\rangle$ are assumed to be far off-resonance. The single-photon detunings are defined as $\Delta_{p}=\omega_{p}-\omega_{21}$ for the probe and $\Delta_{c}=\omega_{c}-\omega_{32}$ for the control field.

To model thermal interacting Rydberg atoms, we define collective transition operators in terms of atomic transition operators distributed in space $\textbf{r}$ and velocity $\textbf{v}$ as
\begin{equation}
\hat{\sigma}_{\alpha\beta}(\textbf{r},\textbf{v},t)=\sum\limits_{j} \hat{\sigma}^{j}_{\alpha\beta}(t)\:\delta(\textbf{r}-\textbf{r}_{j}(t))\: \delta(\textbf{v}-\textbf{v}_{j}(t))\,,
 \label{eq2}
\end{equation}
in analogy to a similar density distribution function developed in~\cite{firstenberg2008,Zhang2014PRA}.
Here, $\hat{\sigma}_{\alpha\beta}^{j}(t)$ is the atomic transition operator for the $j$-th atom  defined as $\hat{\sigma}_{\alpha\beta}^{j}(t)=|\alpha^{j}(t)\rangle\langle\beta^{j}(t)|$, which obeys the Heisenberg equation
\begin{equation}
\frac{d\hat{\sigma}_{\alpha\beta}^{j}(t)}{dt} = \frac{i}{\hbar}\big[H, \hat{\sigma}_{\alpha\beta}^{j}(t)\big] - L\hat{\sigma}_{\alpha\beta}^{j}(t)\;,
\label{heisenberg0}
\end{equation}
where $i$ is the imaginary unit $i=\sqrt{-1}$, $H$ is the Hamiltonian of the system, and $L\hat{\sigma}_{\alpha\beta}^{j}(t)$ denotes the incoherent processes including spontaneous decay and dephasing.
As shown in detail in Appendix \ref{sec:appendix}, the equations of motion of the system can be obtained as~\cite{firstenberg2008}
\begin{align}
\frac{d \hat{\sigma}_{\alpha\beta}(\textbf{r},\textbf{v},t)}{d t} &= \frac{i}{\hbar} [H, \hat{\sigma}_{\alpha\beta}(\textbf{r},\textbf{v},t)] -L\hat{\sigma}_{\alpha\beta}(\textbf{r},\textbf{v},t)\nonumber \\
&\quad - \textbf{v}\cdot\frac{\partial \hat{\sigma}_{\alpha\beta}(\textbf{r},\textbf{v},t)}{\partial \textbf{r}}  \nonumber \\
&\quad -\gamma_{c}\big[\hat{\sigma}_{\alpha\beta}(\textbf{r},\textbf{v},t) - \hat{R}_{\alpha\beta}(\textbf{r},t)F(\textbf{v})\big]\,.
\label{finaleq0}
\end{align}
Here, $F(\textbf{v})=\text{Exp}[-\text{v}^2/\text{v}_{p}^2]/(\sqrt{\pi}\text{v}_{p})^3$ is the Boltzmann distribution function, with $\text{v}_{p} = \sqrt{2k_{B}T/m}$ being the most probable thermal velocity, and  $\gamma_{c}$ is the collision rate. The latter is proportional to the thermal velocity $\gamma_{c}\propto\text{v}_{p}$ and therefore $\gamma_{c}\propto\sqrt{T}$. We further defined the collective transition operator $\hat{R}_{\alpha\beta}(\textbf{r},t)$ as 
\begin{align}
 \hat{R}_{\alpha\beta}(\textbf{r},t) = \int \hat{\sigma}_{\alpha\beta}(\textbf{r},\textbf{v},t) d\textbf{v}\;.
\end{align}
In order to arrive at Eq.~(\ref{finaleq0}) we have assumed that the atomic ensemble has reached to the thermal equilibrium state.
In the right hand side (RHS) of Eq.~(\ref{finaleq0}), the first term describes the internal quantum-mechanical evolution, while the next two terms characterize external classical effects such as thermal motion and atomic collisions, respectively~\cite{firstenberg2008}. 

For convenience, in the following, we use the simplified notations $\hat{\sigma}^{j}_{\alpha\beta}$ and $\hat{\sigma}_{\alpha\beta}$ for  $\hat{\sigma}^{j}_{\alpha\beta}(t)$ and $\hat{\sigma}_{\alpha\beta}(\textbf{r},\textbf{v},t)$. For the setup depicted in Fig.~\ref{fig1}, the Hamiltonian of the $j$-th atom then can be written as~($\hbar=1$)
\begin{align}
 H_{j} &= -\Omega_{p}(\textbf{r}_{j},t)\hat{\sigma}^{j}_{21} -\Omega_{c}(\textbf{r}_{j},t)\hat{\sigma}^{j}_{32} + H.c.\nonumber \\
 &\quad-(\Delta_{p}-\textbf{k}_{p}\cdot\textbf{v}_{j})\hat{\sigma}^{j}_{22} - (\Delta-\Delta\textbf{k}\cdot\textbf{v}_{j})\hat{\sigma}^{j}_{33}\nonumber\\
 &\quad+\sum\limits_{l< j}V_{jl}(t)\hat{\sigma}^{j}_{33}\hat{\sigma}^{l}_{33}\,,
\label{jhamilton0}
\end{align}
where $\Delta=\Delta_{p}-\Delta_{c}$ is the two-photon detuning, and $\Delta\textbf{k}=\textbf{k}_{p}-\textbf{k}_{c}$ is the wavevector difference between the two laser fields. In Eq.~(\ref{jhamilton0}), the last term in the RHS represents the dipole-dipole interaction. We choose a van-der-Waals type potential
\begin{align}
 V_{jl}(t) = \frac{C_{6}}{\big|\textbf{r}_{j}(t)-\textbf{r}_{l}(t)\big|^{6}}\,.
\end{align}
The total Hamiltonian for the thermal Rydberg atoms then is the sum over all single-particle Hamiltonian
\begin{align}
 H = \sum\limits_{j}H_{j}\,.
\end{align}

With Eqs.~(\ref{heisenberg0}) and (\ref{finaleq0}), we can obtain the equations of motion for the relevant transition operators as
\begin{subequations}
\label{eqbasis0}
\begin{align}
 &\frac{\partial\hat{\sigma}^{j}_{12}}{\partial t}=i\big[\Omega_{p}(\textbf{r}_{j})+\Omega_{c}\hat{\sigma}^{j}_{13}+\Delta_{12}(\textbf{v}_{j})\hat{\sigma}^{j}_{12}\big]\,,\\[2mm]
 &\frac{\partial\hat{\sigma}^{j}_{13}}{\partial t}=i\big[\Omega_{c}\hat{\sigma}^{j}_{12}+\Delta_{13}(\textbf{v}_{j})\hat{\sigma}^{j}_{13}\big]-i\sum\limits_{l\neq j}V_{jl}(t)\hat{\sigma}^{j}_{13}\hat{\sigma}^{l}_{33}\,,\\[2mm]
 &\big(\frac{\partial}{\partial t}+\textbf{v}\cdot\frac{\partial}{\partial\textbf{r}}\big)\hat{\sigma}_{12} = i\big[\Delta_{12}(\textbf{v})+i\gamma_{c}\big]\hat{\sigma}_{12} + i\Omega_{p}(\textbf{r})n_{0}F(\textbf{v})\nonumber\\[1mm]
  &\quad\quad\quad\quad\quad\quad\quad\quad\;+i\Omega_{c}\hat{\sigma}_{13} +\gamma_{c}\hat{R}_{12}(\textbf{r},t)F(\textbf{v})\,, \\[2mm]
  &\big(\frac{\partial}{\partial t}+\textbf{v}\cdot\frac{\partial}{\partial\textbf{r}}\big)\hat{\sigma}_{13} = i\big[\Delta_{13}(\textbf{v})+i\gamma_{c}\big]\hat{\sigma}_{13} + i\Omega_{c}\hat{\sigma}_{12}\nonumber\\[2mm]
  &\quad\quad\quad\quad\quad\quad\quad\quad\; - i \sum\limits_{j<l}V_{jl}(t)\hat{\sigma}^{j}_{13}\hat{\sigma}^{l}_{33}\delta(\textbf{r}-\textbf{r}_{j})\delta(\textbf{v}-\textbf{v}_{j})  \nonumber\\[1mm]
  &\quad\quad\quad\quad\quad\quad\quad\quad\; +\gamma_{c}\hat{R}_{13}(\textbf{r},t)F(\textbf{v})\,.
\end{align}
\end{subequations}
Here, $n_{0}=N/V$ is the atomic density, and we have defined 
\begin{subequations}
\begin{align}
\Delta_{12}(\textbf{u})&=\Delta_{p}-\textbf{k}_{p}\cdot\textbf{u}+i\gamma_{12}\,,\\
\Delta_{13}(\textbf{u})&=\Delta-\Delta\textbf{k}\cdot\textbf{u}+i\gamma_{13}\,,
\end{align}
\end{subequations}
with $\textbf{u}\in\{\textbf{v}_{j},\textbf{v}\}$. 

In deriving Eq.~(\ref{eqbasis0}), we have approximated $\langle\hat{\sigma}^{j}_{11}\rangle\simeq 1$, $\langle\hat{\sigma}^{j}_{22}\rangle\simeq 0$, $\langle\hat{\sigma}^{j}_{23}\rangle\simeq 0$, $\langle\hat{\sigma}_{11}\rangle\simeq n_{0}F(\textbf{v})$, $\langle\hat{\sigma}_{22}\rangle\simeq 0$, and $\langle\hat{\sigma}_{23}\rangle\simeq 0$. These approximations are valid in the weak probe limit $\Omega_{p}\ll\Omega_{c}$ as well as in the far-detuned regime $\Delta_{p}\gg\text{k}_{p}\text{v}_{p},\gamma_{12},\Omega_{p}$. Further, we focus on the continuous-wave case where $\Omega_{p}(\textbf{r},t)=\Omega_{p}(\textbf{r})$, $\Omega_{c}(\textbf{r},t)=\Omega_{c}(\textbf{r})$, and assume that the control field is approximately constant across the probe field region,  $\Omega_{c}(\textbf{r})\simeq\Omega_{c}$.

\subsection{\label{Derivation} Nonlocal effects in Rydberg atoms}

In the far-detuned regime, $\hat{\sigma}^{j}_{12}$ can be adiabatically eliminated to derive an equation of motion for the correlation operator $\hat{\sigma}^{j}_{13}\hat{\sigma}^{l}_{33}$ as
\begin{align}
 \frac{\partial(\hat{\sigma}^{j}_{13}\hat{\sigma}^{l}_{33})}{\partial t}
  \simeq&-\frac{i\Omega_{p}(\textbf{r}_{j})\Omega_{c}\hat{\sigma}^{l}_{33}+i\Omega^{2}_{c}\hat{\sigma}^{j}_{13}\hat{\sigma}^{l}_{33}}{\Delta_{12}(\textbf{v}_{j})}\nonumber\\
  &+i\Delta_{13}(\textbf{v}_{j})\hat{\sigma}^{j}_{13}\hat{\sigma}^{l}_{33}-iV_{jl}(t)\hat{\sigma}^{j}_{13}\hat{\sigma}^{l}_{33}\,.
\label{correlation0}
\end{align}
In general, due to the presence of the time-dependent dipole-dipole interaction $V_{jl}(t)$ in thermal Rydberg atoms, it is challenging to solve Eq.~(\ref{correlation0}), even for a steady-state solution. In order to be able to continue with the analytical derivation, we make the model assumption 
\begin{align}
V_{jl}(t)=V_{jl}(0)\,,\label{assumption}
\end{align}
such that the dipole-dipole interaction between different atoms is considered to be time-independent despite the motion of atoms. While there is no formal justification for this assumption, a physical motivation might be that for a dilute atomic ensemble in thermal equilibrium, in a mean-field approximation one could expect that at least the total interaction of one atom with all of its surrounding atoms on average remains constant, i.e., $\sum_{l}V_{jl}(t)/N=V_{jl}(0)$. 
Applying Eq.~(\ref{assumption}), the state-state solution for the correlation operator follows as
\begin{align}
 \hat{\sigma}^{j}_{13}\hat{\sigma}^{l}_{33} = \frac{-\Omega_{c}\Omega_{p}(\textbf{r}_{j})\hat{\sigma}^{l}_{33}}{\Omega_{c}^{2}-\Delta_{12}(\textbf{v}_{j})\Delta_{13}(\textbf{v}_{j})+V_{jl}(0)\Delta_{12}(\textbf{v}_{j})}\,.
\label{correlation1}
\end{align}
Substituting Eq.~(\ref{correlation1}) into Eq.~(\ref{eqbasis0}b) and setting $\partial\hat{\sigma}^{j}_{13}/\partial t=0$, we can obtain the steady-state solution for $\hat{\sigma}^{j}_{13}$, and exploit it to calculate
\begin{align}
 \hat{\sigma}^{j}_{33}&=\hat{\sigma}^{j}_{31}\hat{\sigma}^{j}_{13}\simeq\frac{\Omega_{c}^{2}\big|\Omega_{p}(\textbf{r}_{j})\big|^{2}}{|\Delta_{12}(\textbf{v}_{j})\Delta_{13}(\textbf{v}_{j}) - \Omega_{c}^{2}|^2}\,.
\label{population0}
\end{align}
Here, we have neglected  higher-order contributions of order $O(\Omega_{p}^{4})$. In the paraxial regime, we can assume the spatial slowly-varying envelope approximation (SVEA)
$|\textbf{v}\cdot\partial/\partial\textbf{r}|\ll |\Delta_{12}(\textbf{v})+i\gamma_{c} |, |\Delta_{13}(\textbf{v})+i\gamma_{c}|$.
This allows us to drop the position derivatives in Eq.~(\ref{eqbasis0}c) and (\ref{eqbasis0}d). Using techniques similar to those developed in Ref.~\cite{firstenberg2008,Zhang2014PRA}, we finally obtain the atomic response of the thermal Rydberg atoms as 
\begin{align}
 \rho_{21}(\textbf{r}) =&\langle\hat{R}_{12}(\textbf{r})\rangle \nonumber \\
 =&\Omega_{p}(\textbf{r})\bigg[\frac{M}{D}
 + \frac{i(\gamma_{13}+\gamma_{c})n_{0}^{2}\Omega_{c}^4A}{D}  \nonumber\\
 & 
  \times \int K(\textbf{r}-\textbf{r}')\:\big|\Omega_{p}(\textbf{r}')\big|^{2}\:d\textbf{r}'\bigg]
 \label{thermal0}\,.
\end{align}
The coefficients are defined as 
\begin{subequations}
\label{parameters}
 \begin{align}
 D &= \left[\Delta\text{k}^{2}\text{v}_{p}^{2}-\frac{\Delta\text{k}}{\text{k}_{p}}\Omega_{c}^{2}+\gamma_{13}(\gamma_{13}+\gamma_{c})\right]\frac{i\gamma_{c}G-1}{G} \nonumber\\
 &\quad\;-\left[\frac{\Delta\text{k}}{\text{k}_{p}}(\Delta_{p}+\gamma_{12})+i(\gamma_{13}+\gamma_{c})\right]\Omega_{c}^{2}\,,\\[2mm]
 M &= \left[\Delta\text{k}^{2}\text{v}_{p}^{2}+\gamma_{13}(\gamma_{13}+\gamma_{c})\right]n_{0}\,,\\[2mm]
 G &= \int\frac{F(\textbf{v})\:d^{3}\textbf{v}}{\Delta_{p}-\textbf{k}_{p}\cdot\textbf{v}+i(\gamma_{12}+\gamma_{c})}\,,\\[2mm]
 A &= \int\frac{F(\textbf{v})}{|\Delta_{12}(\textbf{v})\Delta_{13}(\textbf{v}) - \Omega_{c}^{2}|^2}\:d^{3}\textbf{v}\,,\\[2mm]
 K(\textbf{r}) &= \int\frac{\big(1-i\frac{\Delta\textbf{k}\cdot\textbf{v}}{\gamma_{13}+\gamma_{c}}\big)F(\textbf{v})V(\textbf{r})\: d^{3}\textbf{v}}{\Omega_{c}^{2}-\Delta_{12}(\textbf{v})\Delta_{13}(\textbf{v})+V(\textbf{r})\Delta_{12}(\textbf{v})}\,.
 \end{align}
\end{subequations}
Here $G$ and $A$ are related to the Doppler-averaging single- and two- photon spectrum respectively, and $K(\textbf{r})$ corresponds to the dipole-dipole interaction modified by the thermal atomic motion and collisions. Note that in the limit $T\rightarrow 0$, our results reduce to those obtained in Ref.~\cite{Sevincli2011PRL} for ultracold Rydberg atoms, as expected.

\section{\label{dynamics} Propagation dynamics in the paraxial regime}
The propagation dynamics of the probe beam is governed by the Maxwell's equations, which in the paraxial regime can be written as
\begin{align}
\left(\frac{\partial}{\partial\zeta}-\frac{i}{2}\frac{\partial^2}{\partial\xi^2}\right)\Omega_{p}(\xi,\zeta) 
&=i\frac{3\lambda_{p}^2\Gamma_{21}S_{z}}{8\pi}\rho_{21}(\xi,\zeta) 
\label{prop-m0}
\end{align}
We have rescaled to dimensionless coordinates using $\xi=\textbf{r}_{\perp}/S_{t}$, $\zeta=z/S_{z}$ and $S_{z}=\text{k}_{p}S_{t}^2$, with $S_{t}$ and $S_{z}$ being the scales in the transverse and propagation directions, respectively. Since we consider a continuous-wave probe field, the evolution of the probe in the $\zeta$ direction is much slower than that of $K(\xi,\zeta)$. Therefore, in the nonlocal integral Eq.~(\ref{thermal0}),  we can apply the local approximation in $\zeta$-direction as $\Omega_{p}(\xi,\zeta)\simeq\Omega_{p}(\xi)$~\cite{Sevincli2011PRL}. Substituting the atomic response Eq.~(\ref{thermal0}) into Eq.~(\ref{prop-m0}) then results in 
\begin{align}
&\left(\frac{\partial}{\partial\zeta}-\frac{i}{2}\frac{\partial^2}{\partial\xi^2}\right)\psi(\xi,\zeta)
\nonumber \\
&\qquad =i\left [C_{l}+C_{nl} K_{s}(\xi)\ast|\psi(\xi)|^2 \right ] \: \psi(\xi,\zeta)\,,
\label{prop-m1}
\end{align}
where ``$\ast$'' denotes the convolution. We have further set $\Omega_{p}(\xi,\zeta)=\Omega_{p0}\psi(\xi,\zeta)$, where $\psi$ is a normalized shape function satisfying  $|\psi(\xi=\zeta=0)| = 1$. The scaled kernel function $K_{s}(\xi)$ is defined by
\begin{align}
  K_{s}(\xi) = \int\frac{\big(1 - i\frac{\Delta\text{k}\text{v}_{p}v}{\gamma_{13}+\gamma_{c}}\big)e^{-v^2}dvd\eta}{\frac{\Delta_{12}(\text{v})}{\Delta_{p}}+\frac{\Omega_{c}^2 S_{t}^6}{\Delta_{p}C_{6}}[1-\frac{\Delta_{12}(\text{v})\Delta_{13}(\text{v})}{\Omega_{c}^2}](\xi^2+\eta^2)^3},
\end{align}
with $\text{v}=\text{v}_{p}v$ and 
\begin{subequations}
\begin{align}
C_{l}&= \frac{3\lambda_{p}^2\Gamma_{21}S_{z}M}{8\pi D}\,,\\[1ex]
C_{nl}&=\frac{3i\lambda_{p}\Gamma_{21}(\gamma_{13}+\gamma_{c})A\Omega_{c}^{4}S^{5}_{t}n_{0}^{2}\Omega^{2}_{p0}}{4\sqrt{\pi}D\Delta_{p}}\,.
\label{coefficients}
\end{align}
\end{subequations}

\subsection{\label{localdynamics} Local nonlinear propagation dynamics}

\begin{figure}[t]
\includegraphics[width=6.5cm]{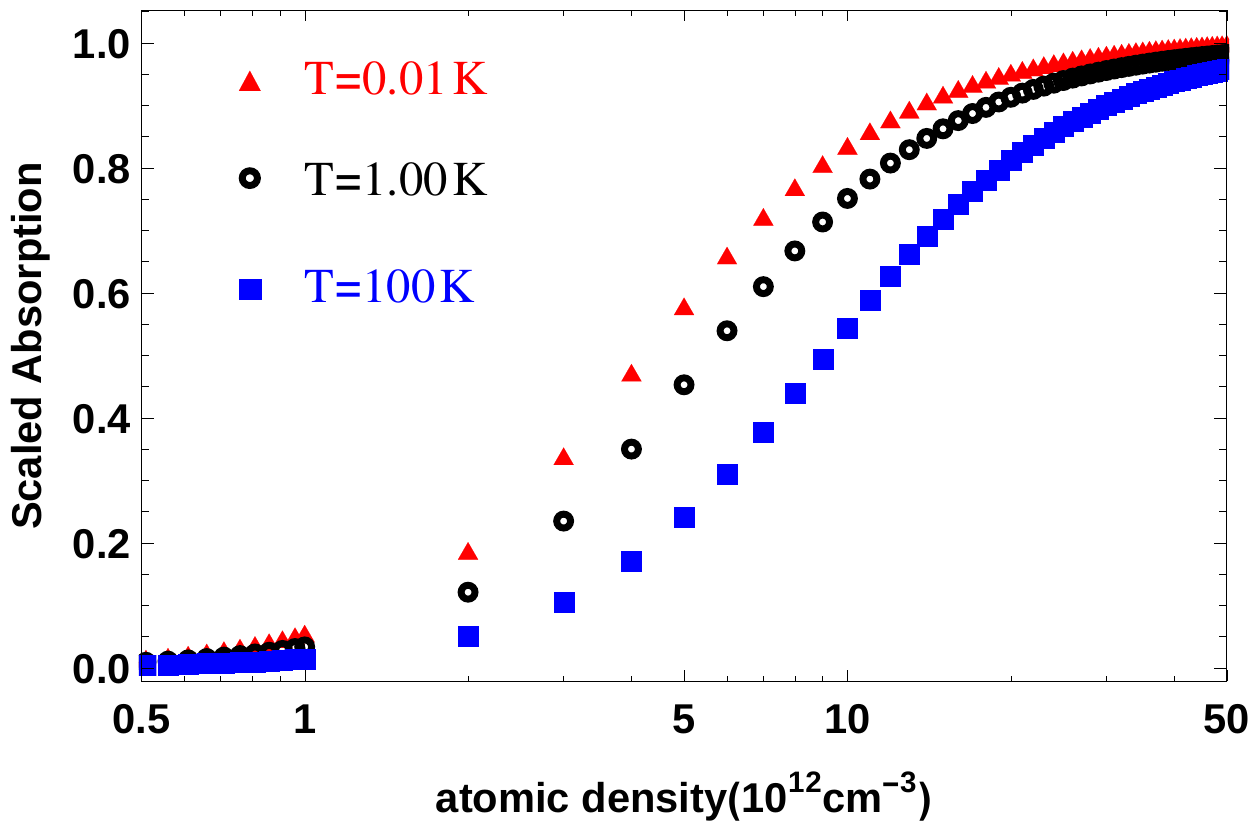}
\caption{(Color online) Scaled absorption against the atomic density at different temperatures. The parameters are chosen as $\lambda_{p}=0.795~\mu\text{m}$, $\lambda_{c}=0.48~\mu\text{m}$, $\Gamma_{21}=2\pi\times 5.75~\text{MHz}$, $\gamma_{12}=\Gamma_{21}/2$, $\gamma_{13}=0.001\gamma_{12}$, $\gamma_{c}=0.01|\Delta\text{k}|\text{v}_{p}$, $\Delta_{p}=\Delta_{c}=0$, $\Omega_{p0}=2\pi\times 0.5~\text{GHz}$, $\Omega_{c}=2\pi\times 10~\text{GHz}$, $C_{6}=140~\text{GHz}~\mu \text{m}^{6}$, and $S_{t}=40~\mu\text{m},z_{L}=1~\text{cm}$.}
\label{fig2}
\end{figure}

So far, our analysis focused on the regime of large single-photon detuning, where the adiabatic elimination is valid. However, in principle, a steady-state solution can be obtained even in the single-photon near-resonant regime. Within the applied approximations, the analytical analysis still holds in this regime, where the nonlinear absorption plays a major role in the propagation dynamics. In this case, the nonlocal effect in Eq.~(\ref{thermal0}) can be neglected which means that $\Omega_{p}(\textbf{r}')=\Omega_{p}(\textbf{r})$~\cite{Sevincli2011PRL}. 
In the absence of dipole-dipole interactions, i.e., $C_{6}=0$, the atomic medium becomes transparent for the probe field because of the EIT configuration. However, in the presence of the interaction, nonlinear absorption is introduced. The thermal atom motion induces a Doppler averaging over all atoms with different velocities. For near-resonant light, we expect Doppler effects  to effectively weaken the nonlinear absorption, since they shift atoms out of resonance with the light  (see also Sec.~\ref{nonlocaldynamics}). 
To analyze the absorption, we numerically solve Eqs.~(\ref{thermal0}) and (\ref{prop-m0}) to obtain the propagation dynamics for a Gaussian-shaped probe field $\Omega_{p}(\xi)=\Omega_{p0}e^{-\xi^2/2}$, from which the scaled absorption against the atomic density $n_{0}$ can be derived. Here, the scaled absorption is defined as the actual absorption divided by the two-level absorption obtained for $\Omega_{c}=0$. 
Results at different temperatures are shown in Fig.~\ref{fig2}. The propagation distance is chosen as $z_{L}=1.0~\text{cm}$, close to one Rayleigh length $z_{R}=1.26~\text{cm}$.  In Fig.~\ref{fig2}, the scaled absorption increases nonlinearly as a function of $n_{0}$, which is in  qualitative agreement with previous calculations~\cite{Petrosyan2011PRL,Garttner2013PRA} and experiments~\cite{Hofmann2013PRL} obtained in ultracold Rydberg gases. Furthermore, as expected, the nonlinear absorption at any specific atomic density is decreased due to Doppler averaging as the temperature rises.

\begin{figure}[t]
\includegraphics[width=7.5cm]{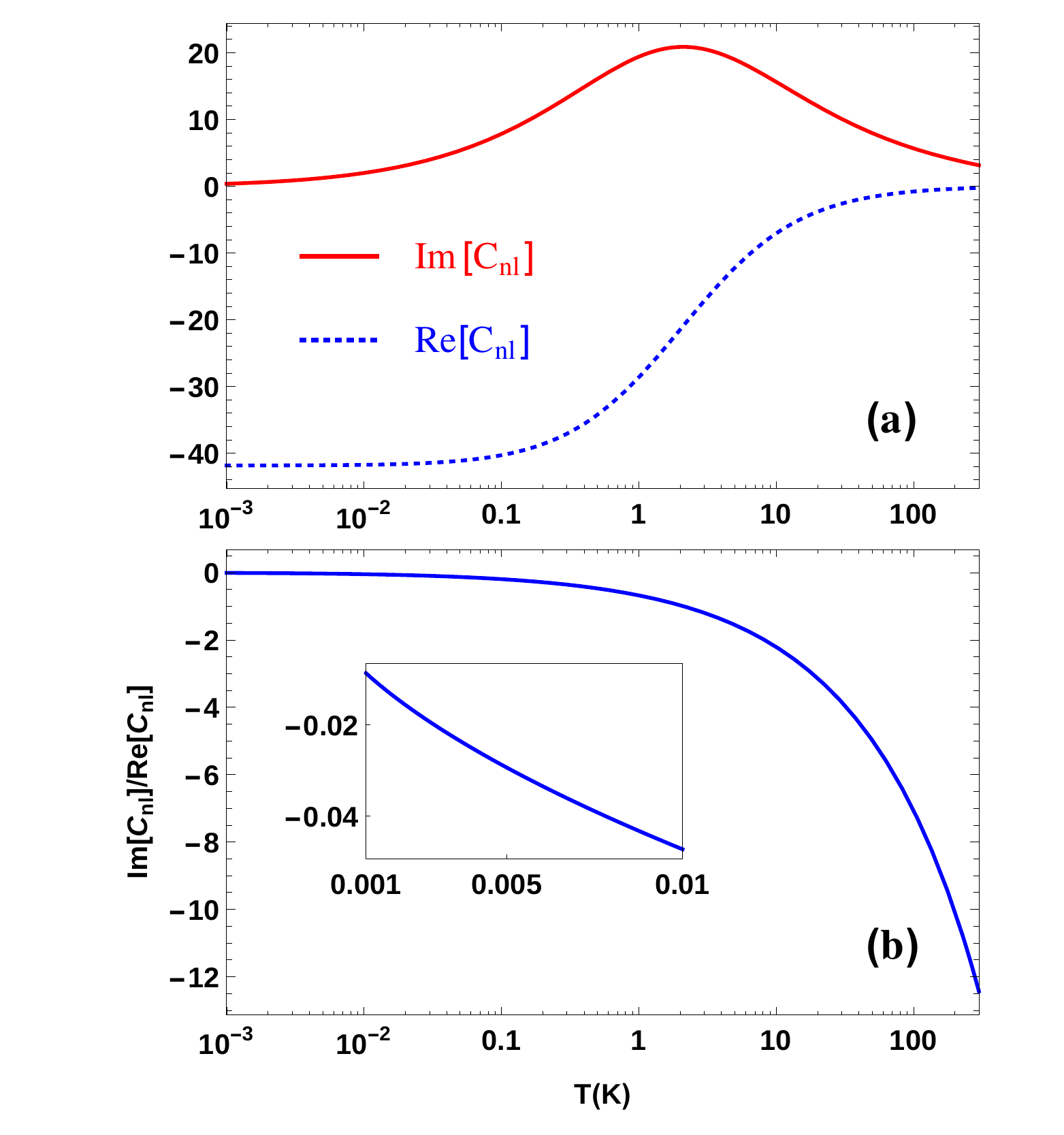}
\caption{(Color online) (a) Nonlinear coefficient $C_{nl}$ and (b) the ratio between its real and imaginary parts $\text{Im}[C_{nl}]/\text{Re}[C_{nl}]$ against temperature $T$. The inset shows the ratio at lower temperatures, where small values indicate that $\text{Im}[C_{nl}]$ can be neglected when $T\rightarrow 0$. Parameters are $n_{0}=1.0\times 10^{20}~\text{m}^{-3}$, $\Delta_{p}=2\pi\times 1.2~\text{GHz}$, $\Delta_{c}=-\Delta_{p}$, and $S_{t}=5R_{c}$ with $R_{c}$ being the blockade radius $R_{c}=(|C_{6}\Delta_{p}|/\Omega_{c}^2)^{1/6}\simeq 0.8~\mu\text{m}$. Other parameters are as in Fig.~\ref{fig2}.}
\label{fig3}
\end{figure}

\subsection{\label{nonlocaldynamics} Nonlocal nonlinear propagation dynamics}

\subsubsection{\label{sec-coefficients}Nonlinear coefficient $C_{nl}$ and scaled nonlocal response $K_{s}(\xi)$}
\begin{figure}[t]
\includegraphics[width=8.5cm]{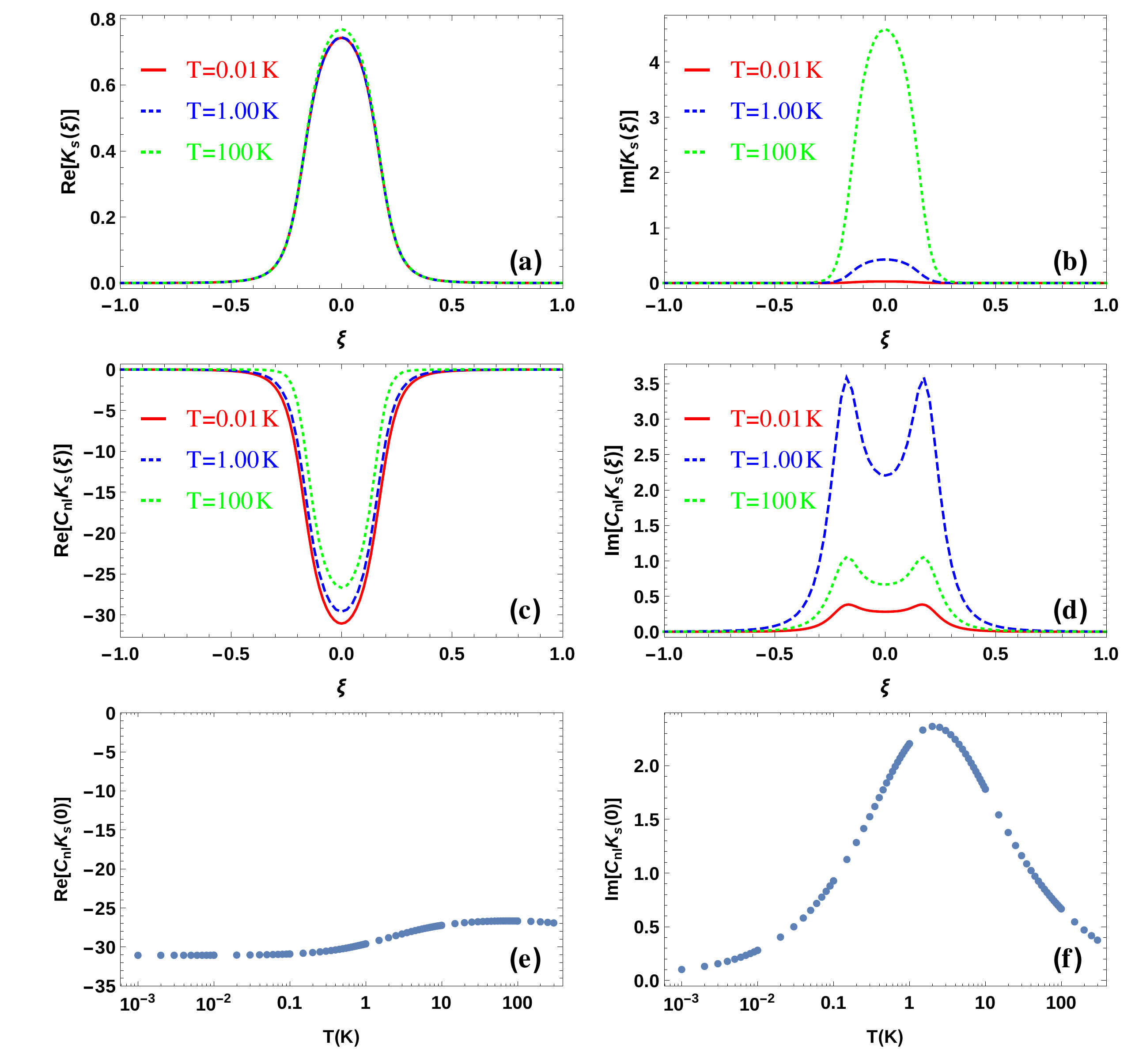}
\caption{(Color online) The real (a) and imaginary (b) parts of the nonlocal response function $K_{s}(\xi)$ and the real (c) and imaginary (d) parts of $C_{nl}K_{s}(\xi)$ at different temperatures. $C_{nl}K_{s}(0)$ as a function of $T$ is plotted in (e) and (f). Parameters are the same as in Fig.~\ref{fig3}.}
\label{fig4}
\end{figure}

Next, we consider the far-detuned regime where the single-photon detunings dominate,  $\Delta_{p}\gg\text{k}_{p}\text{v}_{p}$, $\gamma_{12}$, $\Omega_{p}$. In this regime, the local approximation is only  valid in the propagation direction~\cite{Sevincli2011PRL}, i.e., $\Omega_{p}(\textbf{r}')=\Omega_{p}(\textbf{r}'_{\perp})$ in Eq.~(\ref{thermal0}). The propagation dynamics is mainly characterized by the nonlinear coefficient $C_{nl}$ and the scaled nonlocal response $K_{s}(\xi)$, see Eq.~(\ref{prop-m1}). 
In cold Rydberg atoms, we find $|\text{Im}[C_{nl}]| \ll |\text{Re}[C_{nl}]|$ such that $C_{nl}$ can be approximated as a real number, see Fig.~\ref{fig3}. Then, the real and imaginary parts $\text{Re}[K_{s}(\xi)]$ and $\text{Im}[K_{s}(\xi)]$ of $K_{s}(\xi)$ are directly related to the nonlocal nonlinear dispersion and absorption, respectively. At low temperatures, the nonlocal nonlinear absorption determined by $\text{Im}[K_{s}(\xi)]$ is negligible as compared to the nonlocal nonlinear dispersion, see Fig.~\ref{fig4}(a) and (b). Consequently, the modulational instability (MI)~\cite{Krolikowski2001PRE,Wyller2002PRE} caused by the nonlocal nonlinear dispersion is expected to crucially influence the probe propagation dynamics, as discussed in Ref.~\cite{Sevincli2011PRL}. 

As $T$ increases, we find that $C_{nl}$ and $K_{s}(\xi)$ change in a characteristic way, influencing the propagation dynamics. As shown in Fig.~\ref{fig3}, $\text{Im}[C_{nl}]$ first grows and then decreases, while the absolute value of the negative quantity $\text{Re}[C_{nl}]$ decreases monotonically towards zero. Inspecting the quantities which are defined in Eq.~(\ref{parameters}), we find that this can be mainly attributed to the collision rate $\gamma_{c}$ which is a function of $T$.

Corresponding results for $K_{s}(\xi)$ are shown in Fig.~\ref{fig4}(a) and (b)  for three different temperatures $T\in\{$0.01K, 1K, 100K$\}$. The real part $\text{Re}[K_{s}(\xi)]$ is essentially the same for all three temperatures. However, the imaginary part $\text{Im}[K_{s}(\xi)]$ increases strongly with temperature, while maintaining approximately its shape as a function of $\xi$. This can be understood in a velocity-dependent dressed-state picture. In the far-off resonant regime where $\Delta_{p}>0$, part of the atoms have velocities which Doppler-shift their transition frequency  close to the single-photon resonance. This fraction increases with the mean velocity and thus the temperature, thereby leading to a narrowed and amplified spectrum for the probe after summing over the contributions of all atoms~\cite{zhang2011JPB}. As a result, the amplitude of $K_{s}(\xi)$ for the nonlocal nonlinear  interaction increases with $T$. In contrast, close to the one-photon resonance  $\Delta_{p}\simeq 0$, an 
increasing Doppler shift progressively moves the atom transition frequencies out of resonance with increasing temperature, thus resulting in a broadened and attenuated spectrum for the probe field when summing over the contributions from all atoms with different velocities.  Therefore, in this case, an increasing temperature $T$ reduces the linear and nonlinear absorption for the probe field, as already discussed in Sec.~\ref{localdynamics}.

Finally, we analyze the combined nonlocal contribution $C_{nl}K_{s}(\xi)$, as shown in Figs.~\ref{fig4}(c) and (d). We find that as $T$ increases, $\text{Re}[C_{nl}K_{s}(\xi)]$ remains roughly unchanged. However, $\text{Im}[C_{nl}K_{s}(\xi)]$ first increases with growing temperature, assumes a maximum, and subsequently decreases again in magnitude. To further illustrate this behavior, we show the central value $C_{nl}K_{s}(0)$ against $T$ in Fig.~\ref{fig4}(e) and (f). Again, the maximum of the imaginary part of the nonlocal contribution at intermediate temperatures is clearly visible. By comparing the dependence of $C_{nl}$ and $K_{s}(\xi)$ on the temperature, it is found that this behavior is mainly owning to $C_{nl}$ and therefore related to the temperature-dependent collision rate $\gamma_{c}\propto\sqrt{T}$.  

\subsubsection{Propagation dynamics}
\begin{figure}[t]
\includegraphics[width=8.5cm]{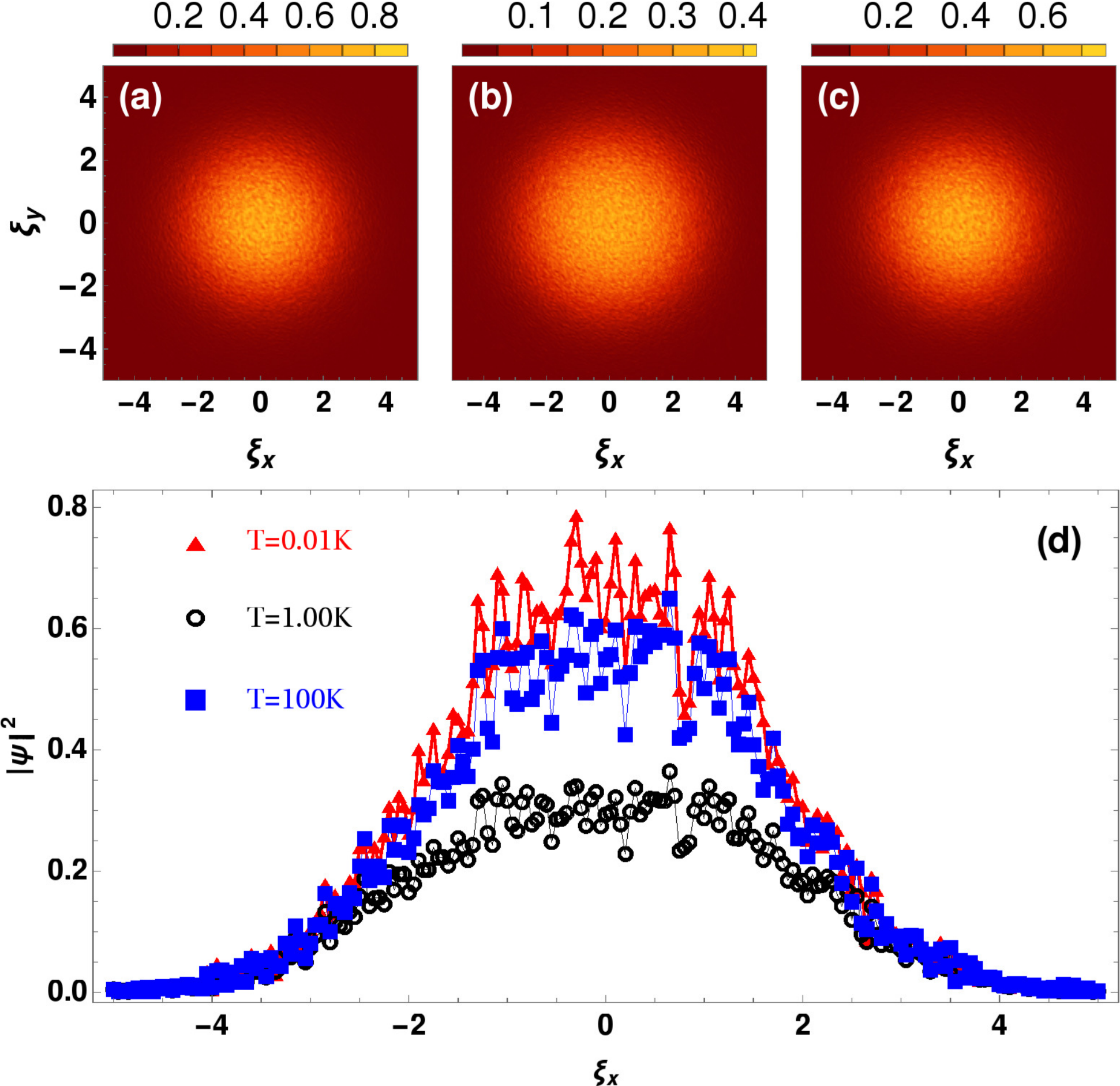}
\caption{(Color online) Probe field propagation through thermal Rydberg atoms at different temperatures $T=0.01K$~(a), $T=1K$~(b), and $T=100K$~(c). (d) shows one-dimensional sections with $\xi_{y}=0$ at the end of the propagation for the three cases (a-c). The localized spikes arise due to the MI effect.  Parameters are  $S_{t}=20R_{c}$, $w_{p}=2$, and $\zeta_{L}=1/16$ which corresponds to $z_{L}\simeq 127.3~\mu \text{m}$.  Other parameters are as in Fig.~\ref{fig3}. 
}
\label{fig5}
\end{figure}

In order to see how the propagation dynamics varies with the temperature, we numerically calculated the probe propagation at different temperatures. The input probe is chosen as a Gaussian
\begin{align}
\psi(\xi_{x},\xi_{y})=e^{-(\xi_{x}^2+\xi_{y}^2)/2w_{p}^2}e^{i\phi(\xi_{x},\xi_{y})}\,,
\end{align}
where the random phase function $\phi(\xi_{x},\xi_{y})$ introduces the spatial noises which may induce MI. Results are shown in Fig.~\ref{fig5}. It can be seen that spatially localized spikes emerge in the output profile of the probe field. As the temperature rises, the output intensity firstly decreases and subsequently grows again, which is consistent with Fig.~(\ref{fig4}). Note that at first sight, this figure might give the impression that the width of the probe output profile increases with the nonlinear absorption. This is an artefact of the different scaling in the plots. Instead, while the absorption changes the beam profile, the width does not increase. 

Owing to the nonlinear absorption introduced by $\text{Im}[C_{nl}K(\xi)]$, a linear stability analysis~\cite{Krolikowski2001PRE,Wyller2002PRE} would not be applicable for the MI effect. The reason is that the plane-wave ansatz assumed in the linear stability analysis is no longer a solution to Eq.~(\ref{prop-m1}) when the absorption is present. In order to understand MI more clearly, we instead define a weighted spatial Fourier spectrum 
\begin{align}
f(\text{k})=\left | \frac{F[\psi(\xi_{x},\xi_{y}=0,\zeta),\xi_{x},\text{k}]}{F[\psi(\xi_{x},\xi_{y}=0,\zeta=0),\xi_{x},\text{k}]}\right | \,,
\end{align}
with $F[\psi,\xi_{x},\text{k}]$ denoting the spatial Fourier transform of $\psi$ from $\xi_{x}$ to $\text{k}$.
This quantity relates the transverse spatial Fourier component $\text{k}$ at propagation distance $\zeta$ to its initial value at $\zeta = 0$, and therefore enables an analysis of potential nonlinear effects throughout the propagation.

\begin{figure}[t!]
\includegraphics[width=6.5cm]{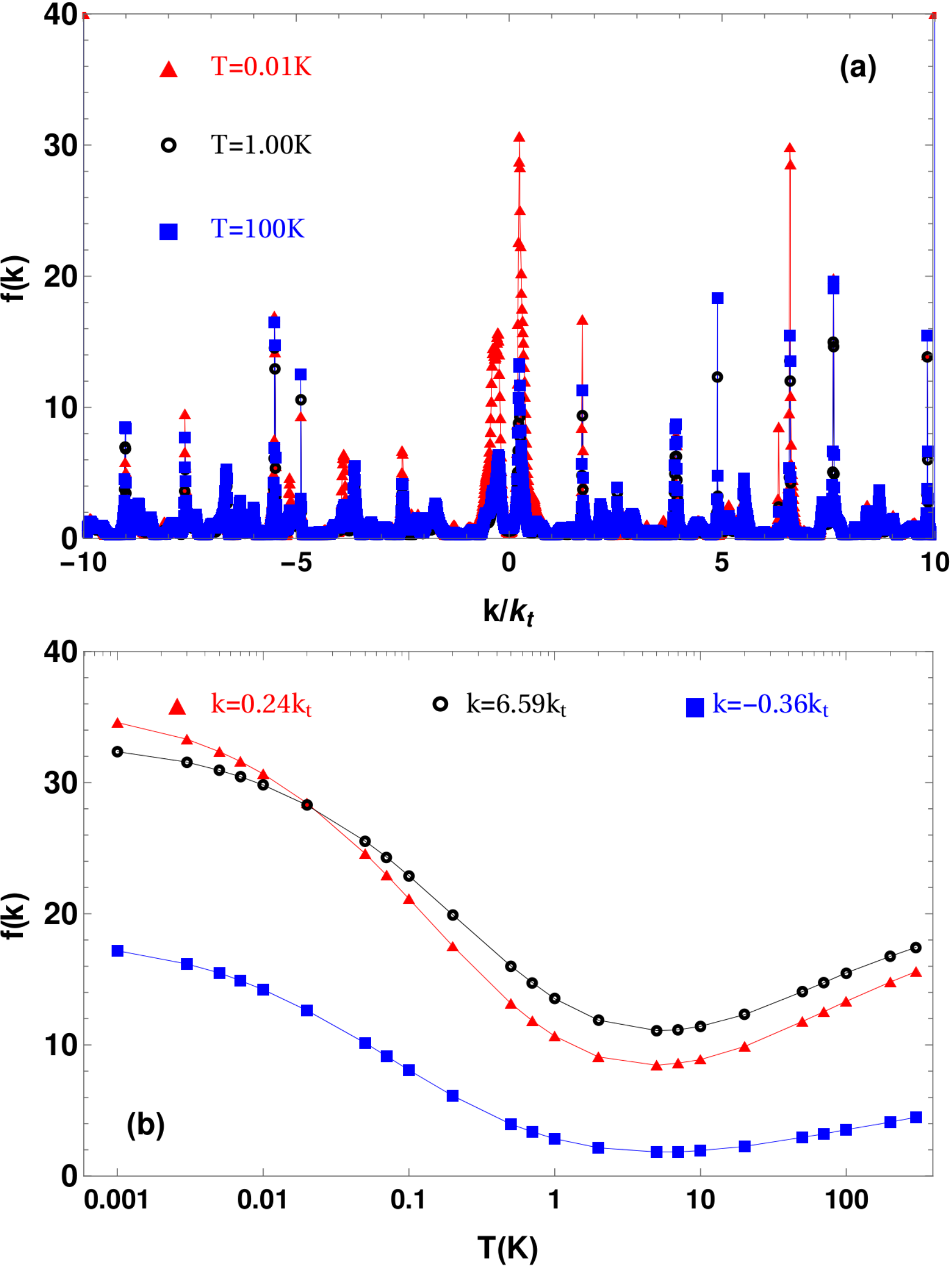}
\caption{(Color online) (a) Weighted Fourier spectrum $f(\text{k})$ of the output probe field at $\xi_{y}=0$. (b) Spectrum as a function of temperature $T$ for three specific wave components $\text{k}\in\{-0.36\text{k}_{t}, 0.24\text{k}_{t}, 6.59\text{k}_{t}\}$. Parameters are as in Fig.~\ref{fig5}, except for $\text{k}_{t}=2\pi/S_{t}$.}
\label{fig6}
\end{figure}

First, we calculate $f(\text{k})$ for three different temperatures. Results are shown in Fig.~\ref{fig6}(a). It can be seen that there are some $\text{k}$ components which are strongly enhanced ($f(\text{k})\gg 1$), while others are not. It should be noted that an enhancement mechanism only becomes visible in this measure if it is strong enough to overcome the nonlinear absorption. Therefore, one could argue that the condition $f(\text{k})>1$ in general is not a necessary condition for the occurrence of MI, as will be discussed in more detail below. 
The random noise $\phi$ is the origin of the asymmetry of the spectrum $f(\text{k})$. Setting $\phi = 0$, we obtained a symmetric spectrum $f(\text{k})$ not shown here.

Next, we analyzed three specific components $\text{k}\in\{-0.36\text{k}_{t}, 0.24\text{k}_{t}, 6.59\text{k}_{t}\}$ in more detail as function of  temperature. These components all feature a large $f(\text{k})\gg 1$ at low temperatures. The results in Fig.~\ref{fig6}(b) show that the weighted Fourier spectrum of these components first decreases, and then increases again, providing a direct signature of the effect of the temperature-dependent nonlocal nonlinear absorption in thermal Rydberg media predicted from the nonlocal coefficient $C_{nl}$ and the scaled nonlocal response $K_{s}(\xi)$ in Sec.~\ref{sec-coefficients}.

\subsubsection{Competition between MI and nonlocal nonlinear absorption}

Owing to the simultaneous presence of MI effect and the nonlocal nonlinear absorption, a competition between these two mechanisms can be expected. Therefore, the weighted spatial wave components $f(\text{k})$ in general do not possess a simple  exponential dependence on the propagation distance. In order to show this, we propagated a two-dimensional plane wave with an additional regular perturbation which has the general form
$\psi(\xi_{x},\zeta=0)= \text{Exp}[i\text{k}_{0}\xi_{x}]+\epsilon_{0}\text{Exp}[i\text{k}_{x}\xi_{x}]$ 
with $\epsilon_{0}$ real and $\epsilon_{0}\ll 1, \text{k}_{0}\neq \text{k}_{x}$. During the propagation, multiple wave components $j(\text{k}_{x}-\text{k}_{0})+\text{k}_{0}$ are generated due to the nonlinear effects in Eq.~(\ref{prop-m1}). We thus make the following ansatz for the propagated function $\psi(\xi,\zeta)$ using a Fourier expansion
\begin{align}
 \psi(\xi_{x},\zeta) = \sum_{j}a_{j}(\zeta)e^{ij(\text{k}_{x}-\text{k}_{0})\xi_{x}}e^{i\text{k}_{0}\xi_{x}}.
 \label{fourier-expansion}
\end{align}
Substituting this form into Eq.~(\ref{prop-m1}) leads to a set of equations for $a_{j}(\zeta)$
\begin{align}
&\left\{\frac{\partial}{\partial\zeta}+\frac{i}{2}[j(\text{k}_{x}-\text{k}_{0})+\text{k}_{0}]^2\right\}a_{j}(\zeta)\nonumber\\[2mm]
 &=iC_{nl}\sum_{p,q}a_{j-p+q}(\zeta)a_{p}(\zeta)a_{q}^{*}(\zeta)K_{F}\left[(p-q)(\text{k}_{x}-\text{k}_{0})\right],
 \label{model}
\end{align}
where $K_{F}[m(\text{k}_{x}-\text{k}_{0})]$ is the Fourier transform of the scaled kernel function $K_{s}(\xi)$ evaluated at $m(\text{k}_{x} - \text{k}_{0})$, i.e.,
\begin{align}
 K_{F}[m(\text{k}_{x}-\text{k}_{0})] = \int^{\infty}_{-\infty}K_{s}(\xi)e^{-im(\text{k}_{x}-\text{k}_{0})\xi}d\xi\,.
\end{align}
$K_{F}[m(\text{k}_{x}-\text{k}_{0})]$ is related to the nonlocal nonlinear response of the medium for the $m$-th spatial frequency component $m(\text{k}_{x}-\text{k}_{0}) + \text{k}_{0}$. In particular, it is worthy to note that $K_{F}(0)=\int^{\infty}_{-\infty} K_{s}(\xi)d\xi$, which means that the nonlocal nonlinear response to $a_{0}$ remains always the same no matter what $\text{k}_{0}$ we choose. Thus, it is not surprising that the nonlocal response to the $m$-th component is determined by its frequency difference from $a_{0}$, i.e., $m(\text{k}_{x}-\text{k}_{0})$, but not by its actual frequency $m(\text{k}_{x}-\text{k}_{0}) + \text{k}_{0}$. 

In general, analytical solutions of this set of equations are challenging. However, over a limited propagation distance $\zeta_{L}$ or for a weak nonlocal response $K_{F}[m(\text{k}_{x}-\text{k}_{0})]$, the set can be safely truncated to $j\leq j_{\text{max}}$. If the propagation distance is sufficiently short, or if $|K_{F}[m(\text{k}_{x}-\text{k}_{0})]|$  decreases sufficiently rapid with order $|m|$, we can limit the analysis to $j_{\text{max}}=1$ and expand the equations up to leading order in $|a_{\pm 1}/a_{0}|$, which gives 
\begin{subequations}
\label{eq-decomposition}
 \begin{align}
 \frac{da_{0}(\zeta)}{d\zeta}=&-i\frac{\text{k}^{2}_{0}}{2} a_{0}(\zeta) + iC_{nl}K_{F}(0)|a_{0}(\zeta)|^2 a_{0}(\zeta)\,,\\ 
 \frac{da_{1}(\zeta)}{d\zeta}=&-\frac{i\text{k}^{2}_{x}}{2}a_{1}(\zeta) +  i C_{nl}\big[K_{F}(0) |a_{0}(\zeta)|^2 a_{1}(\zeta) \nonumber\\
  &+ K_{F}(\text{k}_{x}-\text{k}_{0}) |a_{0}(\zeta)|^2 a_{1}(\zeta) \nonumber \\
  &  + K_{F}(\text{k}_{x}-\text{k}_{0})\,a^{2}_{0}(\zeta) a^{*}_{-1}(\zeta)\big]\,,\\
 \frac{da_{-1}(\zeta)}{d\zeta}=&-\frac{i(2\text{k}_{0}-\text{k}_{x})^2}{2}a_{-1}(\zeta) \nonumber \\
   &+  i C_{nl}\big[K_{F}(0) |a_{0}(\zeta)|^2 a_{-1}(\zeta) \nonumber\\
  &+ K_{F}(\text{k}_{0}-\text{k}_{x}) |a_{0}(\zeta)|^2 a_{-1}(\zeta) \nonumber \\
  &  + K_{F}(\text{k}_{0}-\text{k}_{x})\,a^{2}_{0}(\zeta) a^{*}_{1}(\zeta)\big]\,. 
\end{align}
\end{subequations}
It can be seen from Eq.~(\ref{eq-decomposition}) that $a_{0}$ will experience a self-Kerr phase shift as well as a nonlinear absorption due to the nonlocal nonlinear response $C_{nl}K_{F}(0)$ which is complex. Furthermore, $a_{1}$ and $a_{-1}$ are coupled to each other via $a_{0}$ which can be interpreted as a special type of four-wave mixing process~\cite{agrawal-book}. In general, $K_{F}(\text{k}_{x}-\text{k}_{0})\neq K_{F}(\text{k}_{0}-\text{k}_{x})$, thus $a_{1}(\zeta)$ and $a_{-1}(\zeta)$ can have different evolutions even though $a_{1}(0)=a_{-1}(0)$. For the special case $\text{k}_{0}=0$ considered here, we find that $K_{F}(m\text{k}_{x})=K_{F}(-m\text{k}_{x})$ by exploiting a symmetry of the kernel function $K_{s}(\xi)$ which satisfies $K_{s}(\xi)=K_{s}(-\xi)$. Then Eq.~(\ref{eq-decomposition}) reduces to 
\begin{subequations}
\label{eq-decomposition-special}
 \begin{align}
 \frac{da_{0}(\zeta)}{d\zeta}=&ic_{0}|a_{0}(\zeta)|^2 a_{0}(\zeta)\,, \label{eqa0}\\ 
 \frac{da_{1}(\zeta)}{d\zeta}=&-i\frac{\text{k}^{2}_{x}}{2} a_{1}(\zeta) +  i \big[c_0 |a_{0}(\zeta)|^2 a_{1}(\zeta) \nonumber\\
  &+ c_1 |a_{0}(\zeta)|^2 a_{1}(\zeta) + c_{1}\,a^{2}_{0}(\zeta) a^{*}_{-1}(\zeta)\big]\,, \label{eqa1}\\
  \frac{da_{-1}(\zeta)}{d\zeta}=&-\frac{i\text{k}^{2}_{x}}{2}a_{-1}(\zeta) +  i \big[c_0 |a_{0}(\zeta)|^2 a_{-1}(\zeta) \nonumber\\
  &+ c_1 |a_{0}(\zeta)|^2 a_{-1}(\zeta) + c_{1}\,a^{2}_{0}(\zeta) a^{*}_{1}(\zeta)\big]\,. \label{eqam1}
\end{align}
\end{subequations}
Here, $c_0 = c_0^R + i c_0^I= C_{nl} K_{F}(0)$ and $c_1 =c_1^R + i c_1^I =  C_{nl} K_{F}(\text{k}_x)$, where $c_0^R = \text{Re}[c_0]$ and $c_0^I = \text{Im}[c_0]$ and analogously for $c_1$. 

A further simplification is possible by specializing to an input field of form $\psi(\xi_{x},\zeta=0)= 1 + \epsilon_{0}\text{cos}[\text{k}_{x}\xi_{x}]$. Also in this case, the resulting equations of motion again are Eq.~(\ref{eq-decomposition-special}). This choice imposes initial conditions $a_{1}(0)=a_{-1}(0)$. With this special initial condition, by observing the symmetry of Eqs.~(\ref{eqa1}) and (\ref{eqam1}), we find that $a_{1}(\zeta)=a_{-1}(\zeta)$. Then Eq.~(\ref{eqa1}) can be further simplified to 
\begin{align}
 \frac{da_{1}(\zeta)}{d\zeta}=&-\frac{i\text{k}^{2}_{x}}{2}a_{1}(\zeta) +  i \,c_0\, |a_{0}(\zeta)|^2 a_{1}(\zeta) \nonumber\\
  &+ i\,c_1\, |a_{0}(\zeta)|^2 a_{1}(\zeta)+ i\,c_1\, a^{2}_{0}(\zeta) a^{*}_{1}(\zeta)\,.  \label{a1simplified}
\end{align}
As shown in Appendix~\ref{app-a0}, Eq.~(\ref{eqa0}) can be solved to
\begin{align}
 a_{0}(\zeta) &= 
  \frac{1}{\sqrt{1+2 c_0^I\zeta}} e^{i \frac{c_0^R}{2c_0^I} \ln(1 + 2c_0^I\zeta)}\,.
 \label{a0solution}
\end{align}
We thus find that $a_{0}(\zeta)$ decreases throughout the propagation due to nonlocal nonlinear absorption proportional to $c_0^I$. In the dispersive case where $c_0^I\rightarrow 0$,  Eq.~(\ref{a0solution}) reduces to 
\begin{align}
 a_{0}(\zeta)  \xrightarrow{c_0^I\rightarrow 0} e^{ic_{0}^R\zeta}.
 \label{a0NoAbsorption}
\end{align}
As expected, in this case $a_{0}(\zeta)$ will not experience absorption, but a nonlinear phase modulation throughout the propagation. 

\begin{figure}[t!]
\includegraphics[width=8.5cm]{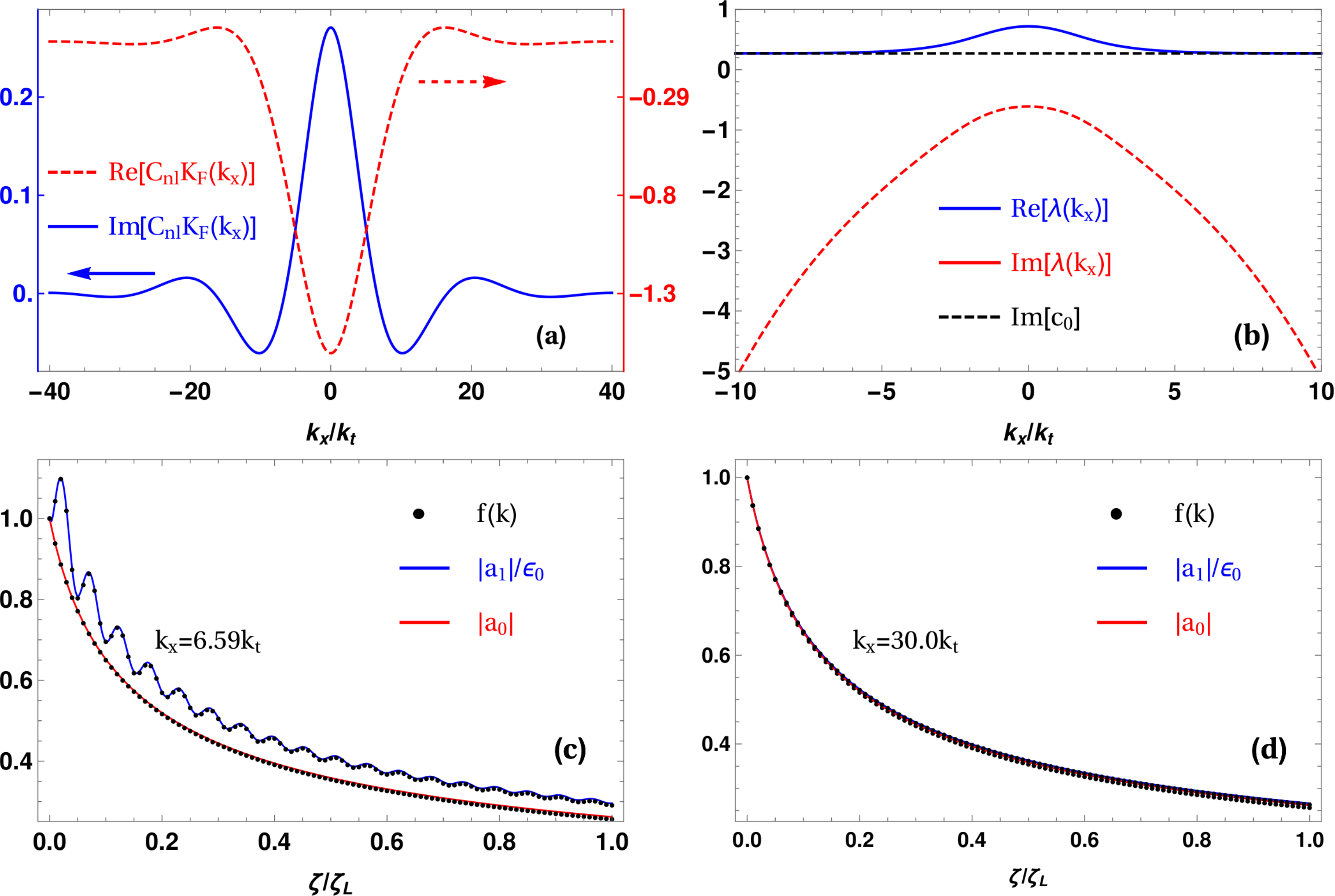}
\caption{(Color online) (a) Nonlocal nonlinear response $C_{nl}K_{F}(\text{k}_{x})$ as a function of $\text{k}_{x}$ in momentum space. (b) shows the related $\lambda(\text{k}_{x})$ defined in Eq.~(\ref{lambdak}). In (c), numerical results $f(\text{k})$ are compared to the analytical solutions $|a_{j}|$ for $\text{k}_{x}=6.59\text{k}_{t}$. As incident probe field, a weakly perturbed plane wave is chosen as defined in the main text. (d) is as (c) except for $\text{k}_{x}=30.0\text{k}_{t}$. Parameters are as in Fig.~\ref{fig6}  but $T=1.0$K.}
\label{fig7}
\end{figure}

Next, we determine $a_1(\zeta)$. Using the solution of $a_{0}(\zeta)$ in Eq.~(\ref{a0solution}), as shown in Appendix~\ref{app-a1}, we find
\begin{align}
\label{a1-expression}
 a_{1}(\zeta) &= \epsilon_{0}\, a_{0}(\zeta) \,e^{-\frac{i\text{k}^{2}_{x}\zeta}{2}}  \left[ s_{1}U\left( \frac{-ic_{1}}{2c_0^I}, \right. \right.
 \nonumber \\
&  \left. 1+\frac{c_1^I}{c_0^I},\frac{i\text{k}^{2}_{x}}{2c_0^I}+i\text{k}^{2}_{x}\zeta\right)  \nonumber \\
 & \left. +s_{2}L \left(\frac{ic_{1}}{2c_0^I},\frac{c_1^I}{c_0^I},\frac{i\text{k}^{2}_{x}}{2c_0^I}+i\text{k}^{2}_{x}\zeta \right ) \right ] \,,
\end{align}
with the confluent hypergeometric function $U(a,b,z)$ and Laguerre polynomials $L(n,a,x) \equiv L^{a}_n(x)$. The coefficients $s_{1}$ and $s_{2}$ are determined by the initial conditions $a_{1}(\zeta=0)=\epsilon_0$ and $da_{1}(\zeta=0)/d\zeta = -\text{k}^{2}_{x}\epsilon_0/2 + i\epsilon_0(c_{0} + 2c_{1})$. To check our analytical solutions given by Eq.~(\ref{a0solution}) and~(\ref{a1-expression}), we compared it to numerical results of Eq.~(\ref{prop-m1}). Fig.~\ref{fig7}(c) and (d) show two examples for $\text{k}_{x}=6.59\text{k}_{t}$ and $\text{k}_{x}=30.0\text{k}_{t}$. We found that our model agrees well with the numerical calculations, and further checks not shown here have also confirmed this agreement for other spatial frequencies $\text{k}_x$. 

Due to the involved confluent hypergeometric function and generalized Laguerre polynomial contribution, it is hard to interpret the physics underlying  $a_{1}(\zeta)$ in an analytical manner. We therefore proceed by approximating the solution of $a_1(\zeta)$ for small propagation distances $\delta\sim 0$, enabling one to analyze the evolution from any starting point $\zeta_0$ to $\zeta = \zeta_0 + \delta$. We find
\begin{align}
 a_{1}(\zeta) = &\epsilon_{0}(1 + 2c_0^I\zeta)^{\frac{i(c_{0}+c_{1})}{2c_0^I}}\nonumber\\[2mm]
                &\times e^{-(c_0^I + ic_1^R)\zeta}\big[s_{1}e^{-\lambda(\text{k}_{x})\zeta}+s_{2}e^{\lambda(\text{k}_{x})\zeta}\big] \\
&= \epsilon_0 a_0(\zeta) (1 + 2c_0^I\zeta)^{\frac{ic_{1}}{2c_0^I}}\, e^{-(c_0^I + ic_1^R)\zeta} \nonumber\\
& \quad \times\big[s_{1}e^{-\lambda(\text{k}_{x})\zeta} + s_{2}e^{\lambda(\text{k}_{x})\zeta}\big]\,,
 \label{a1_solution_approx}
\end{align}
where $s_1$ and $s_2$ depend on $\zeta_0$, and 
\begin{align}
 \lambda(\text{k}_{x}) = \frac{1}{2}\sqrt{4|c_{1}|^2+(2c_0^I-i\text{k}^2_{x} + 2ic_1^R)^2}\,.
 \label{lambdak}
\end{align}
From Eq.~(\ref{a1_solution_approx}) we conclude that $\lambda(\text{k}_{x})$ is associated to the MI in absorptive nonlocal nonlinear media. In the purely dispersive case where $\text{Im}[C_{nl}K_{F}(\text{k}_{x})]=0$, MI occurs only for select $\text{k}$ wave components. Interestingly, when nonlinear absorption is present, $\lambda(\text{k}_{x})$ is complex for all $\text{k}_{x}$, see Eq.~(\ref{lambdak}) and Fig.~\ref{fig7}(b). Thus exponential gain contributions may contribute at all spatial frequencies. However, the nonlinear absorption $c_0^I$ competes with this gain contribution.
Note that in local nonlinear media where only dispersive effects enter, i.e., $K_{F}(\text{k}_{x})=1$ and $C_{nl}=C^{*}_{nl}$, the exponential index $\lambda(\text{k}_{x})$  reduces to 
\begin{align}
 \lambda(\text{k}_{x}) = \frac{i}{2}\sqrt{\text{k}^2_{x}(\text{k}^2_{x}-4C_{nl})}\,,
\end{align}
which agrees with the result obtained in Refs.~\cite{Krolikowski2001PRE,Wyller2002PRE}.

\subsubsection{\label{4wm}Effect of the four-wave mixing process}
To further interpret the origin of the exponential growth components, in this subsection~\ref{4wm}, we  artificially set $c_{1}=0$, which physically means that the four-wave mixing process is neglected. Then, one finds
\begin{align}
 a_{1}(\zeta) = \epsilon_0 (1 + 2c^{I}_{0}\zeta)^{\frac{c_{0}}{2c^{I}_{0}}}e^{-i\text{k}^{2}_{x}\zeta}=\epsilon_{0}a_{0}(\zeta)e^{-i\text{k}^{2}_{x}\zeta}\,.
 \label{a1-no-fwm}
\end{align}
As a consequence, $|a_{1}(\zeta)|$ decays monotonically due to the nonlocal nonlinear absorption of $a_0$ denoted by $c^{I}_{0}$. Since the coupling to $a_{-1}(\zeta)$ is neglected,  the evolution of $a_{1}(\zeta)$ is similar to that of $a_{0}(\zeta)$, except for a different amplitude and an additional phase. 
This result can be furthered tested in our system. By calculating the nonlocal nonlinear response $C_{nl}K_{F}(\text{k}_{x})$ in momentum space as shown in Fig.~\ref{fig7}(a), we find that $c_{1}=C_{nl}K_{F}(\text{k}_{x})$ gradually reduces to zero as $|\text{k}_{x}|\rightarrow\infty$. We thus can choose a large $\text{k}_{x}~(=30.0\text{k}_{t})$ which has $c_{1}\simeq 0$ in order to approximately realize the considered parameter case. The corresponding propagating dynamics is shown in Fig.~\ref{fig7}(d). As expected, $a_{1}(\zeta)$ decays in the same way as $a_{0}(\zeta)$.
We thus conclude that in the absence of the four-wave mixing process, the higher order spatial Fourier components $a_{\pm 1}$ do not exhibit exponential growth contributions.

\begin{figure}[t!]
\includegraphics[width=8.5cm]{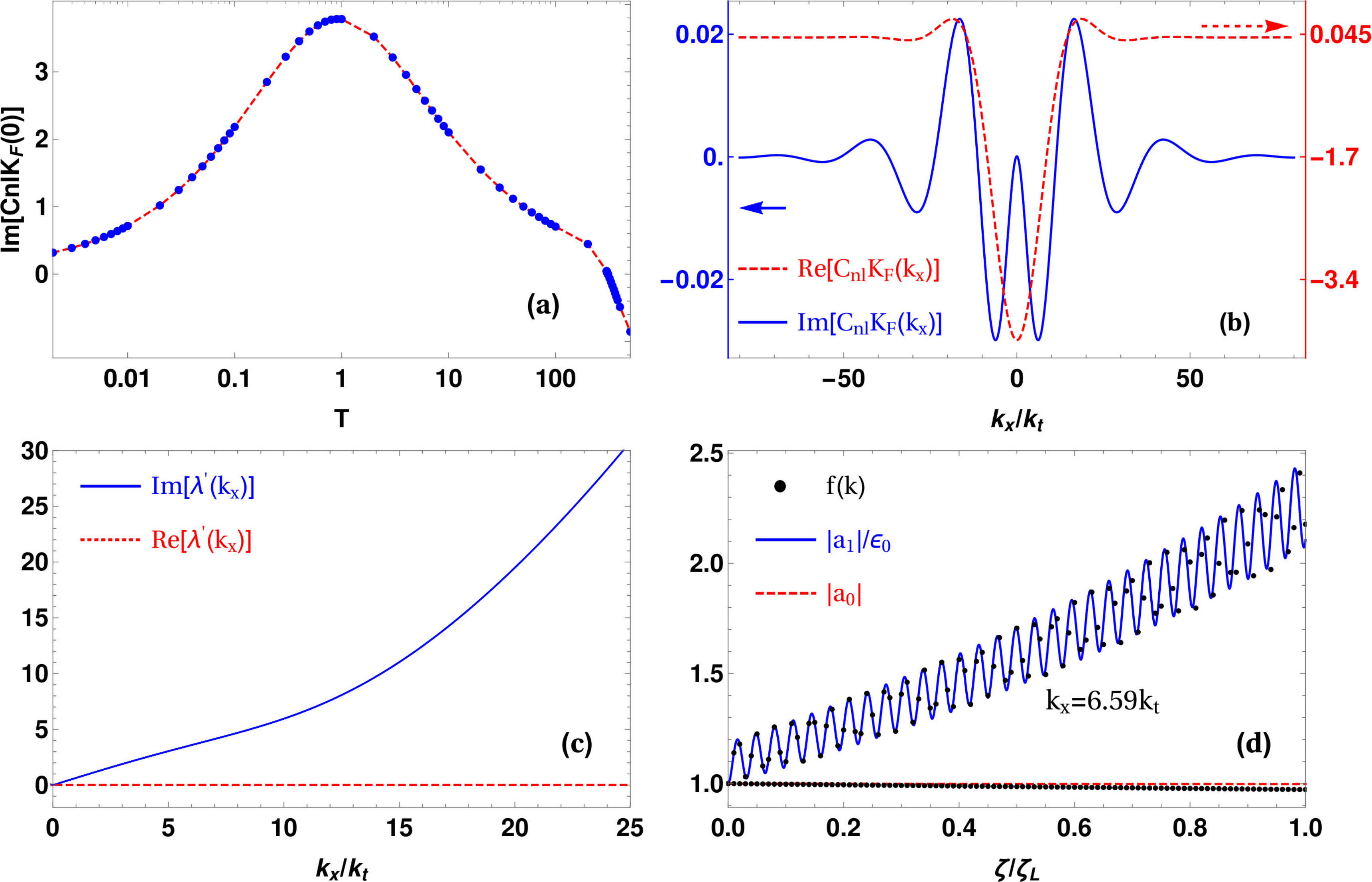}
\caption{(Color online) (a) $\text{Im}[c_{0}]=\text{Im}[C_{nl}K_{F}(0)]$ as a function of temperature $T$. It can be seen that $\text{Im}[c_{0}]\approx 0$ at $T=308.5K$. For this temperature, the corresponding nonlocal nonlinear response $C_{nl}K_{F}(\text{k}_{x})$ in momentum space and $\lambda^{'}(\text{k}_{x})$ defined in Eq.~(\ref{lambdak-case1}) are shown in (b) and (c), respectively. (d) compares $f(\text{k})$ and $|a_{j}|$ for $\text{k}_{x}=6.59\text{k}_{t}$. Parameters are  $\Delta_{p}=2\pi\times 0.8~\text{GHz}$, $\Omega_{p0}=2\pi\times 0.25~\text{GHz}$, $\Omega_{c}=2\pi\times 5~\text{GHz}$; Other parameters are chosen as in Fig.~\ref{fig3}.}
\label{fig8}
\end{figure}

\subsubsection{Effect of pump depletion}
One might argue that the exponential growth due to MI could be restricted to short propagation distances $\delta\sim 0$ where Eq.~(\ref{a1_solution_approx}) is valid, such that the competition between MI and nonlocal nonlinear absorption would occur only around $\zeta\sim 0$. In order to analyze this in more detail, we for the moment manually set the nonlinear absorption $c^{I}_{0}=\text{Im}[c_{0}]=0$.
Then, Eq.~(\ref{a0solution}) reduces to $a_{0}(\zeta)=\text{Exp}[ic^{R}_{0}\zeta]$, such that this case can be interpreted as ``undepleted pump approximation'', if $a_0$ is considered a pump field for the higher order spatial Fourier components $a_{\pm 1}$. As shown in Appendix~\ref{app-a1-case1}, the solution of $a_{1}(\zeta)$ becomes
\begin{align}
 a_{1}(\zeta) =\epsilon_0 e^{ic^{R}_{0}\zeta-c^{I}_{1}\zeta}\bigg[s^{'}_{1}e^{\lambda^{'}(\text{k}_{x})\zeta} + s^{'}_{2}e^{-\lambda^{'}(\text{k}_{x})\zeta}\bigg]\,,
 \label{a1-no-absorption}
\end{align}
with
\begin{align}
  \lambda^{'}(\text{k}_{x}) = \frac{1}{2}\sqrt{4|c_{1}|^2-(\text{k}^2_{x} - 2c_1^R)^2}\,.
  \label{lambdak-case1}
\end{align}
For $\text{k}^2_{x}<4c^{R}_{1}$, we find that $\lambda^{'}(\text{k}_{x})>c^{I}_{1}$. Thus, in the undepleted pump case, $a_{1}(\zeta)$ grows exponentially despite the presence of the nonlocal nonlinear absorption represented by $c^{I}_{1}$. Interestingly, this condition coincides with the condition for the occurrence of MI  in the purely dispersive case~\cite{Krolikowski2001PRE,Wyller2002PRE}.

In order to check the prediction given by Eq.~(\ref{a1-no-absorption}) without manually setting $c^I_{0} = 0$, we have to first identify suitable parameters which satisfy $\text{Im}[c_{0}]=0$. For this, we study $c_{0}$ as function of temperature $T$, as shown in Fig.~\ref{fig8}(a). For the chosen laser field parameters, $\text{Im}[c_{0}] $  vanishes for approximately $T=308.5K$. The corresponding $C_{nl}K_{F}(\text{k}_{x})$ and $\lambda^{'}(\text{k}_{x})$ are shown in Fig.~\ref{fig8}(b) and \ref{fig8}(c). Note that only positive $k_x$ values are shown since $\lambda^{'}(\text{k}_{x})$ is symmetric with respect to $\text{k}_{x}$. 
We find that our model predicts that most $\text{k}_{x}$ components will be enhanced, indicated by negative $\text{Im}[C_{nl}K_{F}(\text{k}_{x})]$. This prediction could serve as a validity check for our model assumption Eq.~(\ref{assumption}) for the chosen parameters. From Fig.~\ref{fig8}(d), we  find that Eq.~(\ref{a1-no-absorption}) agrees well to the numerical results. This figure further shows that the weak amplification of $a_{1}(\zeta)$ with increasing $\zeta$ is due to a negative $\text{Im}[c_{1}]$, and the oscillating behavior arises  due to the interference between the two parts in Eq.~(\ref{a1-no-absorption}) where $\lambda^{'}(\text{k}_{x}=6.59\text{k}_{t})$ is purely imaginary. This observation also helps in understanding the oscillation already observed in Fig.~\ref{fig7}(c).

\begin{figure}[t!]
\includegraphics[width=8.5cm]{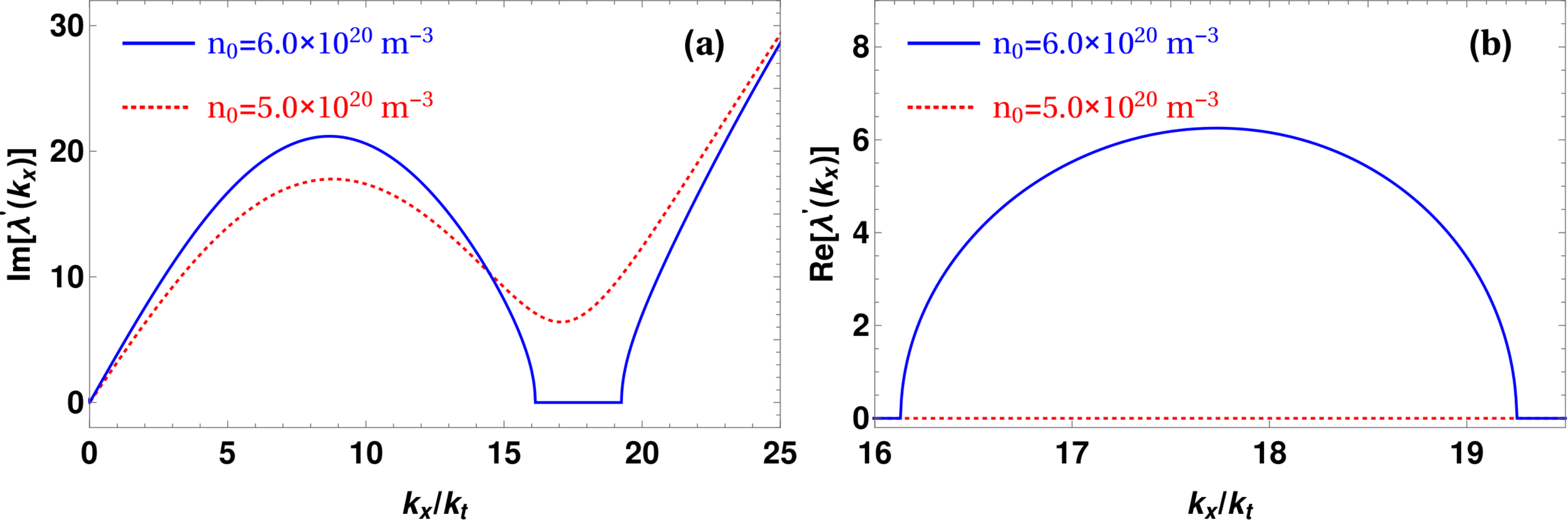}
\caption{(Color online) Modulational instability at higher atomic density. (a) and (b) show the real and imaginary parts of $\lambda^{'}(\text{k}_{x})$ for two different atomic densities, respectively. At $n_{0}=6.0\times 10^{20}\text{m}^{-3}$, a range
of $\text{k}_{x}$ having real $\lambda^{'}(\text{k}_{x})$ appears, which relates to MI. Parameters are as in Fig.~\ref{fig8}.}
\label{fig9}
\end{figure}

\begin{figure}[t!]
\includegraphics[width=8.6cm]{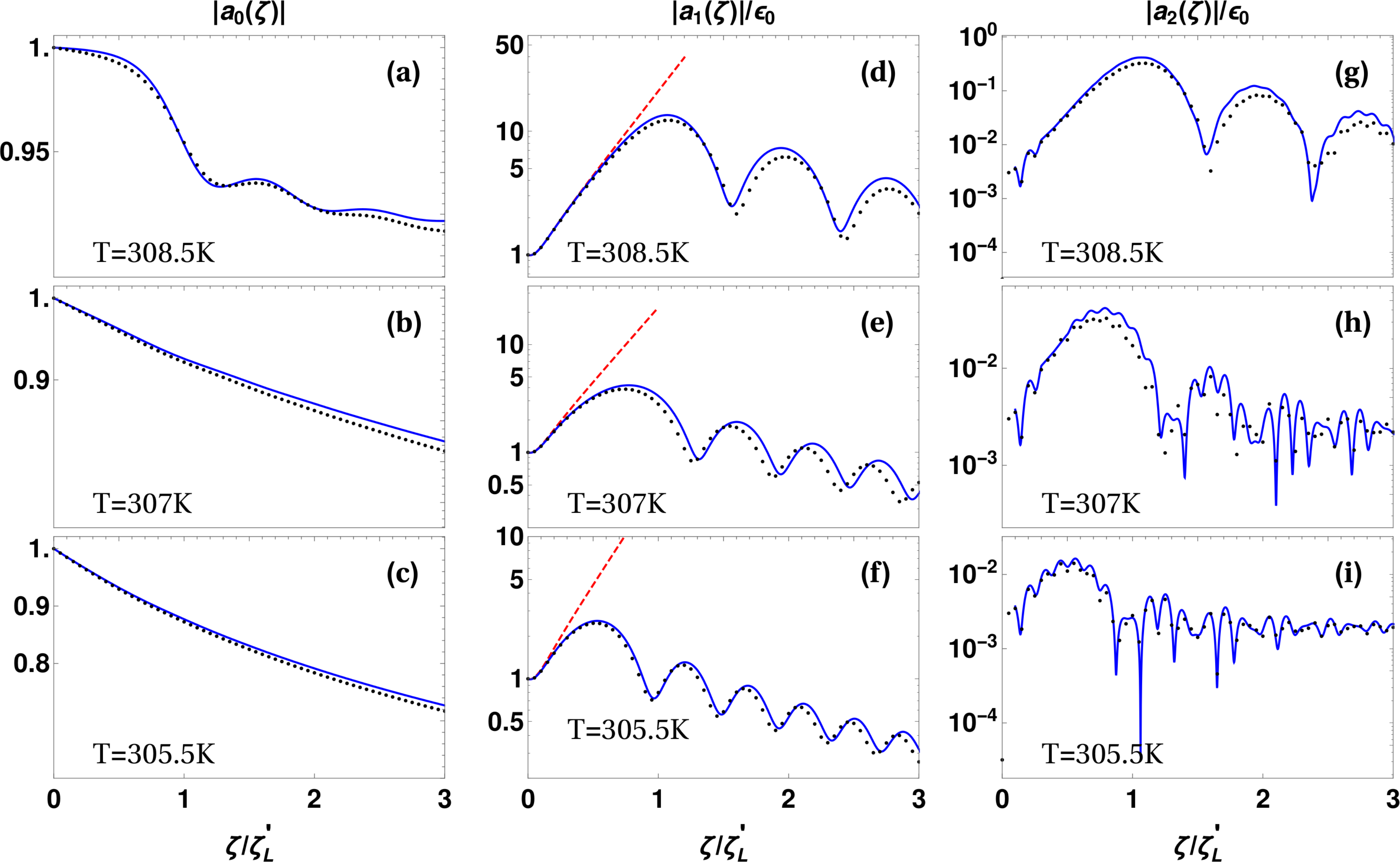}
\caption{(Color online) Evolution of Fourier components $|a_{0}(\zeta)|$, $|a_{1}(\zeta)|$ and $|a_{2}(\zeta)|$ against propagation distance. The black dots denote $f(\text{k})$, solid blue lines represent the result from a full numerical solution to Eq.~(\ref{model}), and the red solid lines in the middle column show Eq.~(\ref{a1-no-absorption}). Note that the middle and right columns are plotted in logarithmic scale. Parameters are $\text{k}_{x}=16.5\text{k}_{t}$, $\zeta^{'}_{L}=\zeta_{L}/25$, and other parameters are as in Fig.~\ref{fig9}.}
\label{fig10}
\end{figure}

Note that due to the high temperature, the nonlocal nonlinear response is weakened such that $\lambda^{'}(\text{k}_{x})$ becomes purely imaginary for the parameters in Fig.~\ref{fig8}. Real $\lambda^{'}(\text{k}_{x})$, which may give rise to MI, can be obtained by adjusting the parameters, for example, increasing the atomic density. This is shown in Fig.~\ref{fig9}, in which $\lambda^{'}(\text{k}_{x})$ becomes real in a certain range of $\text{k}_{x}$ at atomic density $n_{0}=6.0\times 10^{20}\text{m}^{-3}$. To investigate this further we analyze the case $\text{k}_{x}=16.5\text{k}_{t}$ in Fig.~\ref{fig10}. Subpanel (d) shows that $a_{1}(\zeta)$ grows exponentially over a short range of propagation distance. When propagating further, the coupled dynamics between different modes gives rise to suppression of MI and turns the propagation dynamics from exponential growth to regular oscillation, which leads to a discrepancy between the predictions of Eq.~(\ref{a1-no-absorption}) (red solid lines) and the 
numerical results (blue dashed lines).

To analyze the effect of weak nonlocal nonlinear absorption $\text{Im}[c_{0}]$ on the MI effect, we calculate the propagation dynamics of the same wave component $\text{k}_{x}=16.5\text{k}_{t}$ at lower temperatures which introduce gradually stronger nonlocal nonlinear absorption for $a_{0}$. The results are shown in Fig.~\ref{fig10}(e) and (f). While initial exponential growth is still present, its slope is reduced. Furthermore, the suppression of MI occurs at shorter propagation distances for higher absorption $\text{Im}[c_{0}]$. Note that this deviation between numerical calculations and the results of Eq.~(\ref{a1-no-absorption}) coincides with the breakdown of the ``undepleted pump approximation'' for $a_{0}$ assumed in the derivation of Eq.~(\ref{a1-no-absorption}), as can be seen in the left column of Fig.~\ref{fig10}.

Similar results can also be obtained, e.g., for the laser parameters chosen in Fig.~\ref{fig3}. However, we found that in this case, vanishing $\text{Im}[c_{0}]$ requires high temperatures around  $T=699K$ and densities $n_{0}=1.1\times 10^{21}~\text{m}^{-3}$, which are more challenging to realize in practice. Moreover, from further numerical calculations, we found that the MI effect can be obtained at lower temperatures and atomic densities than in Fig.~\ref{fig10} if the probe detuning $\Delta_{p}$ and the control field Rabi frequency $\Omega_{c}$ are further reduced. This is, however, limited by our model assumptions that the probe field should be far-detuned and that $\Omega_{p}\ll\Omega_{c}$.
\section{\label{summary} Discussions and Conclusions}

We have developed a model to describe thermal Rydberg atoms interacting with two laser fields in EIT configuration. Our results lead us to the conclusion that MI may be observed in thermal Rydberg gases over short propagation distances. But in contrast to the cold atom case, in the initial stage of the propagation, there is a competition between the nonlocal nonlinear absorption and MI due to the nonlocal nonlinear dispersion. For longer propagation distances, the coupled dynamics of the Fourier modes $a_{i}$ prevents the occurrence of exponential growth indicative of MI. This turnover to the coupled dynamics is also connected to the breakdown of the undepleted pump approximation, in which the magnitude of the initial Fourier mode $a_0$ remains constant.

\begin{figure}[t!]
\includegraphics[width=7cm]{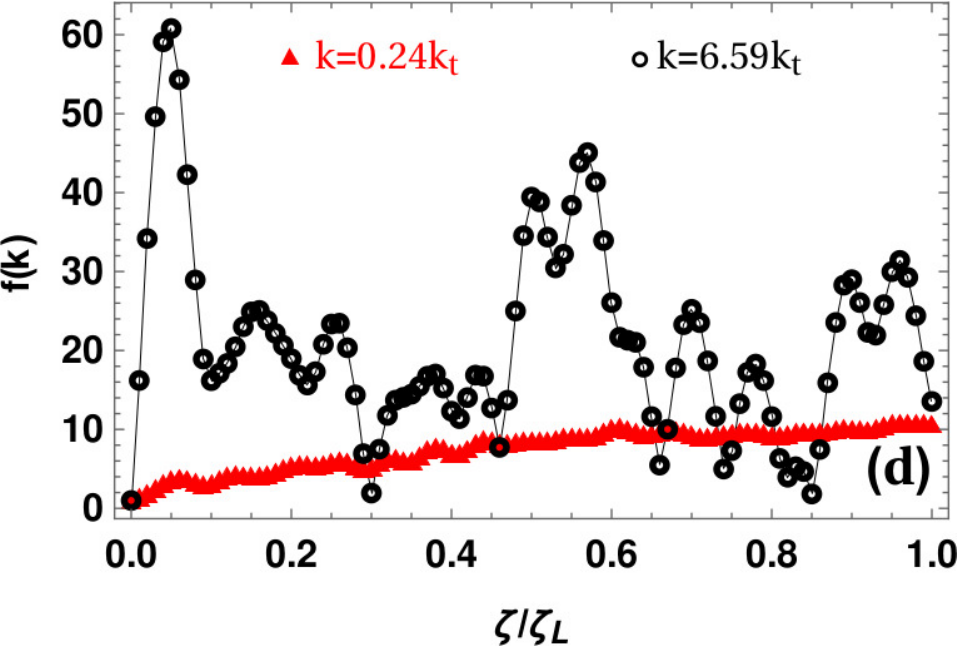}
\caption{(Color online) Weighted Fourier spectrum $f(\text{k})$ against the propagation distance for two modes $\text{k}_{x}=0.24\text{k}_{t}$ and $\text{k}_{x}=6.59\text{k}_{t}$. The initial probe field is a Gaussian beam as previously employed for Fig.~\ref{fig5}. Parameters are  as in Fig.~\ref{fig5}.}
\label{fig11}
\end{figure}

However, the calculations are based on the crucial approximation that the time variation in the interatomic coupling $V(t)$ is neglected despite the atom motion. This approximation enabled us to obtain analytical results, but is likely to limit the validity of the present model. In turn, our results could serve as a test for this mean field like approximation in thermal Rydberg gases. A further theoretical investigation of this approximation is challenging, but should be a next step to improve the reliability of the theoretical predictions.

For more complicated input probe field profiles comprising many $k$ wave components, the regular oscillations of the Fourier components observed in our calculations may not been seen. The reason is that from Eq.~(\ref{prop-m1}), we find that in general all wave components interact with each other such that the resulting dependence on propagation distance for a specific component becomes more intricate and unpredictable. As an example, Fig.~\ref{fig11} shows two of the wave components against the propagation distance for the same input probe profile previously used in Fig.~\ref{fig5}. It is found that the amplification of component $\text{k}_{x}=6.59\text{k}_{t}$ against the propagation distance is irregular and different from that in the simple case shown in Fig.~\ref{fig7}(c). However, we found in our numerical calculations that wave component with smaller $\text{k}_{x}$ evolve more regularly, see the result for $\text{k}_{x}=0.24\text{k}_{t}$ in Fig.~\ref{fig11}. One reason for this could be that 
components with small $\text{k}_{x}$ have a larger initial amplitude in the Gaussian probe field, and are further associated to a stronger initial MI growth rate  [see Fig.~\ref{fig7}(b)], which could reduce the effect of other Fourier components.

Triggered by the invalidity of the linear analysis of MI in absorptive nonlocal nonlinear media, we further developed a generalized model to analyze the MI effect in absorptive nonlinear systems based on the idea of Fourier decomposition. Independent from the first model aiming at describing thermal Rydberg atoms, it enables us to capture MI in absorptive nonlocal nonlinear systems in general. Here we have focused on cases in which the Fourier space could be truncated to few modes. It remains a task for the future to obtain analytical solutions when also higher frequency components have to be taken into account.

We thank Martin G\"{a}rttner for fruitful discussions. We are grateful for funding by the German Science Foundation (DFG, Sachbeihilfe EV 157/2-1).

\appendix

\numberwithin{equation}{section}

\section{\label{sec:appendix}Theoretical model}

In this Appendix, we provide a detailed description of our model for the thermal Rydberg atoms, following the method in~\cite{firstenberg2008,Zhang2014PRA}. All quantities not defined here are already given in the main text.

We start from the collective transition operator $\hat{\sigma}_{\alpha\beta}(\textbf{r},\textbf{v},t)$ introduced in Eq.~(\ref{eq2}). From the Heisenberg equation Eq.~(\ref{heisenberg0}), we find
\begin{align}
\frac{d \hat{\sigma}_{\alpha\beta}(\textbf{r},\textbf{v},t)}{d t} &=\sum\limits_{j} \frac{d\hat{\sigma}^{j}_{\alpha\beta}(t)}{dt}\:\delta(\textbf{r}-\textbf{r}_{j}(t))\: \delta(\textbf{v}-\textbf{v}_{j}(t))\nonumber \\
&\quad +\sum\limits_{j} \hat{\sigma}^{j}_{\alpha\beta}(t)\:\frac{\partial\delta(\textbf{r}-\textbf{r}_{j}(t))}{\partial t}\: \delta(\textbf{v}-\textbf{v}_{j}(t))  \nonumber \\
&\quad +\sum\limits_{j} \hat{\sigma}^{j}_{\alpha\beta}(t)\:\delta(\textbf{r}-\textbf{r}_{j}(t))\: \frac{\partial\delta(\textbf{v}-\textbf{v}_{j}(t))}{\partial t}\,.
\label{inieq}
\end{align}
For the second term in the RHS of this equation, we have
\begin{align}
 &\sum\limits_{j} \hat{\sigma}^{j}_{\alpha\beta}(t)\:\frac{\partial\delta(\textbf{r}-\textbf{r}_{j}(t))}{\partial t}\: \delta(\textbf{v}-\textbf{v}_{j}(t))\nonumber \\
 & = \sum\limits_{j} \hat{\sigma}^{j}_{\alpha\beta}(t)\:\frac{\partial\textbf{r}_{j}(t)}{\partial t}\frac{\partial\delta(\textbf{r}-\textbf{r}_{j}(t))}{\partial\textbf{r}_{j}}\: \delta(\textbf{v}-\textbf{v}_{j}(t)) \nonumber \\
 & = -\textbf{v}\cdot\frac{\partial\hat{\sigma}_{\alpha\beta}(\textbf{r},\textbf{v},t)}{\partial\textbf{r}}\,.
\end{align}
Assuming that the atomic system is relaxed to the thermal equilibrium state, the last term describing atomic collision in the RHS of Eq.~\ref{inieq} can be rewritten as
\begin{align}
&\sum\limits_{j} \hat{\sigma}^{j}_{\alpha\beta}(t)\:\delta(\textbf{r}-\textbf{r}_{j}(t))\: \frac{\partial\delta(\textbf{v}-\textbf{v}_{j}(t))}{\partial t}\nonumber\\
&=\gamma_{c}\big[\hat{\sigma}_{\alpha\beta}(\textbf{r},\textbf{v},t) - \hat{R}_{\alpha\beta}(\textbf{r},t)F(\textbf{v})\big]\,.
\end{align}
with all quantities defined in the main text. This finally leads to Eq.~\ref{finaleq0} in the main text.

In order to obtain the atomic response of the thermal Rydberg gas to the incident probe field, we  derive the equations of motions
\begin{subequations}
 \label{singleeq}
 \begin{align}
 &\frac{\partial\hat{\sigma}^{j}_{12}}{\partial t}=i\big[\Omega_{p}(\textbf{r}_{j},t)(\hat{\sigma}^{j}_{11}-\hat{\sigma}^{j}_{22})+\Omega_{c}(\textbf{r}_{j},t)\hat{\sigma}^{j}_{13}+\Delta_{12}(\textbf{v}_{j})\hat{\sigma}^{j}_{12}\big],\\[1mm]
 &\frac{\partial\hat{\sigma}^{j}_{13}}{\partial t}=i\big[-\Omega_{p}(\textbf{r}_{j},t)\hat{\sigma}^{j}_{23}+\Omega_{c}(\textbf{r}_{j},t)\hat{\sigma}^{j}_{12}+\Delta_{13}(\textbf{v}_{j})\hat{\sigma}^{j}_{13}\big] \nonumber\\
 &\quad\quad\quad-i\sum\limits_{l\neq j}V_{jl}(t)\hat{\sigma}^{j}_{13}\hat{\sigma}^{l}_{33}\,,
 \end{align}
\end{subequations}
with $\Delta_{12}(\textbf{v}_{j})$ and $\Delta_{13}(\textbf{v}_{j})$ defined in the main text. Similarly, for the collective transition operators, we find
\begin{subequations}
 \begin{align}
  \big(\frac{\partial}{\partial t}+\textbf{v}\cdot\frac{\partial}{\partial\textbf{r}}\big)\hat{\sigma}_{12} &= i\big[\Delta_{12}(\textbf{v})+i\gamma_{c}\big]\hat{\sigma}_{12} + i\Omega_{p}(\textbf{r},t)(\hat{\sigma}_{11}-\nonumber\\[1mm]
  &\quad \hat{\sigma}_{22})+i\Omega_{c}(\textbf{r},t)\hat{\sigma}_{13} +\gamma_{c}\hat{R}_{12}(\textbf{r},t)F(\textbf{v}), \\[1mm]
  \big(\frac{\partial}{\partial t}+\textbf{v}\cdot\frac{\partial}{\partial\textbf{r}}\big)\hat{\sigma}_{13} &= i\big[\Delta_{13}(\textbf{v})+i\gamma_{c}\big]\hat{\sigma}_{13} + i\Omega_{c}(\textbf{r},t)\hat{\sigma}_{12}\nonumber\\[2mm]
  &\quad -i\Omega_{p}(\textbf{r},t)\hat{\sigma}_{23}+\gamma_{c}\hat{R}_{13}(\textbf{r},t)F(\textbf{v}) \nonumber\\[1mm]
  &\quad - i \sum\limits_{j<l}V_{jl}(t)\hat{\sigma}^{j}_{13}\hat{\sigma}^{l}_{33}\delta(\textbf{r}-\textbf{r}_{j})\delta(\textbf{v}-\textbf{v}_{j}).
 \end{align}
\label{collectiveeq}
\end{subequations}
In general, Eqs.~(\ref{collectiveeq}) are difficult to solve due to the dipole-dipole interaction terms, even in the ``frozen gas'' limit $V_{jl}(t)=V_{jl}(t=0)$. For a first step, here we limit ourselves to the far-detuned regime where $\Delta_{p},\Delta_{c}\gg\text{k}_{p}\text{v}_{p},\Omega_{p},\gamma_{12},\gamma_{13},\gamma_{c}$ and also the weak probe limit $\Omega_{c}\gg\Omega_{p}$, where it was shown that an analytical solution in steady state is possible at $T=0$~\cite{Sevincli2011PRL}. In this far-detuned regime, we  approximate  $\langle\hat{\sigma}^{j}_{11}\rangle\simeq 1,\langle\hat{\sigma}^{j}_{22}\rangle\simeq 0,\langle\hat{\sigma}^{j}_{23}\rangle\simeq 0$, and find
\begin{subequations}
\begin{align}
 \langle\hat{\sigma}_{11}\rangle &=\sum\limits_{j}\langle\hat{\sigma}^{j}_{11}\rangle\delta(\textbf{r}-\textbf{r}_{j})\delta(\textbf{v}-\textbf{v}_{j}) 
 =n_{0}F({\textbf{v}}) \,,\\
 \langle\hat{\sigma}_{22}\rangle &=0\,,\\
 \langle\hat{\sigma}_{23}\rangle &=0\,,
\end{align}
\end{subequations}
where we have applied the relation
\begin{align}
\sum\limits_{j}f(\textbf{r}_{j},\textbf{v}_{j})
&=\sum\limits_{j,N\rightarrow\infty}\frac{N}{V}F({\textbf{v}_{j}})f(\textbf{r}_{j},\textbf{v}_{j})\frac{V}{N}\Delta\textbf{v}_{j}\nonumber\\
&=n_{0}\int f(\textbf{r}_{j},\textbf{v}_{j})F({\textbf{v}_{j}})d^{3}\textbf{r}_{j}d^{3}\textbf{v}_{j}\,.
\label{sumtoint}
\end{align}
Next we assume the continuous-wave case $\Omega_{p}(\textbf{r},t)=\Omega_{p}(\textbf{r})$ and $\Omega_{c}(\textbf{r},t)=\Omega_{c}(\textbf{r})$, and that the beam diameter of $\Omega_{c}(\textbf{r})$ is much larger than that of $\Omega_{p}(\textbf{r})$ such that $\Omega_{c}(\textbf{r})\simeq\Omega_{c}$ and $\Omega_{c}$ can be chosen as real. With these considerations, Eq.~(\ref{singleeq}) and (\ref{collectiveeq}) can be simplified to
\begin{subequations}
\begin{align}
 &\frac{\partial\hat{\sigma}^{j}_{12}}{\partial t}=i\big[\Omega_{p}(\textbf{r}_{j})+\Omega_{c}\hat{\sigma}^{j}_{13}+\Delta_{12}(\textbf{v}_{j})\hat{\sigma}^{j}_{12}\big],\\[2mm]
 &\frac{\partial\hat{\sigma}^{j}_{13}}{\partial t}=i\big[\Omega_{c}\hat{\sigma}^{j}_{12}+\Delta_{13}(\textbf{v}_{j})\hat{\sigma}^{j}_{13}\big]-i\sum\limits_{l\neq j}V_{jl}(t)\hat{\sigma}^{j}_{13}\hat{\sigma}^{l}_{33} \label{eqbasisb},\\[2mm]
 &\big(\frac{\partial}{\partial t}+\textbf{v}\cdot\frac{\partial}{\partial\textbf{r}}\big)\hat{\sigma}_{12} = i\big[\Delta_{12}(\textbf{v})+i\gamma_{c}\big]\hat{\sigma}_{12} + i\Omega_{p}(\textbf{r})n_{0}F(\textbf{v})\nonumber\\[1mm]
  &\quad\quad\quad\quad\quad\quad\quad\quad\;+i\Omega_{c}\hat{\sigma}_{13} +\gamma_{c}\hat{R}_{12}(\textbf{r},t)F(\textbf{v}) \label{eqbasisc},\\[2mm]
  &\big(\frac{\partial}{\partial t}+\textbf{v}\cdot\frac{\partial}{\partial\textbf{r}}\big)\hat{\sigma}_{13} = i\big[\Delta_{13}(\textbf{v})+i\gamma_{c}\big]\hat{\sigma}_{13} + i\Omega_{c}\hat{\sigma}_{12}\nonumber\\[2mm]
  &\quad\quad\quad\quad\quad\quad\quad\quad\; - i \sum\limits_{j<l}V_{jl}(t)\hat{\sigma}^{j}_{13}\hat{\sigma}^{l}_{33}\delta(\textbf{r}-\textbf{r}_{j})\delta(\textbf{v}-\textbf{v}_{j})  \nonumber\\[1mm]
  &\quad\quad\quad\quad\quad\quad\quad\quad\; +\gamma_{c}\hat{R}_{13}(\textbf{r},t)F(\textbf{v}).
 \end{align}
 \label{eqbasis}
 \end{subequations}
In the far-detuned regime, $\hat{\sigma}^{j}_{12}$ can be adiabatically eliminated, leading to
\begin{align}
 \hat{\sigma}^{j}_{12}=-\frac{\Omega_{p}(\textbf{r}_{j})+\Omega_{c}\hat{\sigma}^{j}_{13}}{\Delta_{12}(\textbf{v}_{j})}.
\label{rho12}
 \end{align}
Plugging Eq.~(\ref{rho12}) into Eq.~(\ref{eqbasisb}), we find
\begin{align}
 \frac{\partial\hat{\sigma}^{j}_{33}}{\partial t} &=\frac{\partial\hat{\sigma}^{j}_{31}}{\partial t} \hat{\sigma}^{j}_{13}+  \hat{\sigma}^{j}_{31}\frac{\partial\hat{\sigma}^{j}_{13}}{\partial t}\nonumber\\
 &\simeq -2\gamma_{13}\hat{\sigma}^{j}_{33}.
\end{align}
For a Rydberg atom, the lifetime of the highly-lying Rydberg state $|3\rangle$ is much longer than that of the intermediate state $|2\rangle$, which means $\gamma_{13}\ll\gamma_{12},\Omega_{c}$. Thus, $\hat{\sigma}^{j}_{33}$ decays much slower than the other terms and can be treated as constant. With this approximation, we find
\begin{align}
 \frac{\partial(\hat{\sigma}^{j}_{13}\hat{\sigma}^{l}_{33})}{\partial t} =&\frac{\partial\hat{\sigma}^{j}_{13}}{\partial t} \hat{\sigma}^{l}_{33}+\hat{\sigma}^{j}_{13}\frac{\partial\hat{\sigma}^{l}_{33}}{\partial t}
  \simeq \frac{\partial\hat{\sigma}^{j}_{13}}{\partial t} \hat{\sigma}^{l}_{33}\nonumber\\
  \simeq&-\frac{i\Omega_{p}(\textbf{r}_{j})\Omega_{c}\hat{\sigma}^{l}_{33}+i\Omega^{2}_{c}\hat{\sigma}^{j}_{13}\hat{\sigma}^{l}_{33}}{\Delta_{12}(\textbf{v}_{j})}\nonumber\\
&  +i\Delta_{13}(\textbf{v}_{j})\hat{\sigma}^{j}_{13}\hat{\sigma}^{l}_{33}-iV_{jl}(t)\hat{\sigma}^{j}_{13}\hat{\sigma}^{l}_{33},
\label{correlation}
\end{align}
where we have neglected the higher order terms $\sum\nolimits_{m\neq j,l}V_{jm}(t)\hat{\sigma}^{j}_{13}\hat{\sigma}^{m}_{33}\hat{\sigma}^{l}_{33}$ which are of order $O(\Omega_{p}^5)$.
Making use of the key assumption of the model
\begin{align}
V_{jl}(t)=V_{jl}(0)[=V_{jl}(t=0)],
\end{align}
discussed in the main text, a steady-state solution for $\hat{\sigma}^{j}_{13}\hat{\sigma}^{l}_{33}$ can be obtained from Eq.~(\ref{correlation}), 
\begin{align}
 \hat{\sigma}^{j}_{13}\hat{\sigma}^{l}_{33} = \frac{-\Omega_{c}\Omega_{p}(\textbf{r}_{j})\hat{\sigma}^{l}_{33}}{\Omega_{c}^{2}-\Delta_{12}(\textbf{v}_{j})\Delta_{13}(\textbf{v}_{j})+V_{jl}(0)\Delta_{12}(\textbf{v}_{j})}\,.
\end{align}
Setting $\partial\hat{\sigma}^{j}_{13}/\partial t=0$, we have
\begin{align}
 \hat{\sigma}^{j}_{13}&=\frac{\Omega_{c}\Omega_{p}(\textbf{r}_{j})}{\Delta_{12}(\textbf{v}_{j})\Delta_{13}(\textbf{v}_{j}) - \Omega_{c}^{2}}\bigg(1 -\Delta_{12}(\textbf{v}_{j})  \nonumber\\
 & \times \sum\limits_{l\neq j}\frac{V_{jl}(0)\hat{\sigma}^{l}_{33}}{\Omega_{c}^{2}-\Delta_{12}(\textbf{v}_{j})\Delta_{13}(\textbf{v}_{j})+V_{jl}(0)\Delta_{12}(\textbf{v}_{j})}\bigg)\,,
\end{align}
%
and thus
\begin{align}
 \hat{\sigma}^{j}_{33}&=\hat{\sigma}^{j}_{31}\hat{\sigma}^{j}_{13}\simeq\frac{\Omega_{c}^{2}\big|\Omega_{p}(\textbf{r}_{j})\big|^{2}}{|\Delta_{12}(\textbf{v}_{j})\Delta_{13}(\textbf{v}_{j}) - \Omega_{c}^{2}|^2}\,.
\label{population}
\end{align}
In the derivation of Eq.~(\ref{population}), we have also neglected the higher-order contributions $O(\Omega_{p}^4)$. Substituting Eq.~(\ref{correlation}) and (\ref{population}) back into Eq.~(\ref{eqbasis}) results in
\begin{align}
   &\textbf{v}\cdot\frac{\partial\hat{\sigma}_{13}}{\partial\textbf{r}} \nonumber \\
   &= i\big[\Delta_{13}(\textbf{v})+i\gamma_{c}\big]\hat{\sigma}_{13} + i\Omega_{c}\hat{\sigma}_{12}
   +\gamma_{c}\hat{R}_{13}(\textbf{r},t)F(\textbf{v})\nonumber\\
  &+i\sum\limits_{j}\sum\limits_{l\neq j}\frac{\Omega_{c}\Omega_{p}(\textbf{r}_{j})V_{jl}(0)\delta(\textbf{r}-\textbf{r}_{j})\delta(\textbf{v}-\textbf{v}_{j})}{\Omega_{c}^{2}-\Delta_{12}(\textbf{v}_{j})\Delta_{13}(\textbf{v}_{j})+V_{jl}(0)\Delta_{12}(\textbf{v}_{j})}\nonumber \\
& \qquad  \times \frac{\Omega_{c}^{2}\big|\Omega_{p}(\textbf{r}_{l})\big|^{2}}{|\Delta_{12}(\textbf{v}_{l})\Delta_{13}(\textbf{v}_{l}) - \Omega_{c}^{2}|^2} \,.
\end{align}
Applying Eq.~(\ref{sumtoint}) to this equation, one finds
\begin{align}
 & \textbf{v}\cdot\frac{\partial\hat{\sigma}_{13}}{\partial\textbf{r}} \nonumber\\
 &= i\big[\Delta_{13}(\textbf{v})+i\gamma_{c}\big]\hat{\sigma}_{13} + i\Omega_{c}\hat{\sigma}_{12}
   +\gamma_{c}\hat{R}_{13}(\textbf{r},t)F(\textbf{v})\nonumber\\
  &\quad+i\,n_{0}^{2}\,\Omega_{c}^{3}\,A\,F(\textbf{v})\Omega_{p}(\textbf{r}) \nonumber \\
  & \quad \times \int\frac{V(\textbf{r}-\textbf{r}')\big|\Omega_{p}(\textbf{r}')\big|^{2}d^{3}\textbf{r}'}{\Omega_{c}^{2}-\Delta_{12}(\textbf{v})\Delta_{13}(\textbf{v})+V(\textbf{r}-\textbf{r}')\Delta_{12}(\textbf{v})}   
\label{nonlocal}
\end{align}
where $V(0)$ is replaced by $V(\textbf{r}-\textbf{r}')$, and
\begin{align}
 A= \int\frac{F(\textbf{v})}{|\Delta_{12}(\textbf{v})\Delta_{13}(\textbf{v}) - \Omega_{c}^{2}|^2}d^{3}\textbf{v}\,.
\end{align}
Setting $\partial \hat{\sigma}^{j}_{12}/\partial t=0$ in Eq.~(\ref{eqbasisc}), the steady-state solution for $\hat{\sigma}_{12}$ becomes
\begin{align}
  &\textbf{v}\cdot\frac{\partial\hat{\sigma}_{12} }{\partial\textbf{r}}= i\big[\Delta_{12}(\textbf{v})+i\gamma_{c}\big]\hat{\sigma}_{12} + i\Omega_{p}(\textbf{r})n_{0}F(\textbf{v}) \nonumber \\
  & \quad +i\Omega_{c}\hat{\sigma}_{13} +\gamma_{c}\hat{R}_{12}(\textbf{r},t)F(\textbf{v})\,.
  \label{sigma12}
\end{align}

The propagation dynamics of the probe field is now completely determined by Eqs.~(\ref{nonlocal}), (\ref{sigma12}) and the propagation equation. Moreover, in the paraxial regime, where the slowly-varying envelope approximation (SVEA) in the spatial dimension is valid,  we have
\begin{align}
 \bigg|\textbf{v}\cdot\frac{\partial}{\partial\textbf{r}}\bigg|\ll\big|\Delta_{12}(\textbf{v})+i\gamma_{c}\big|,\big|\Delta_{13}(\textbf{v})+i\gamma_{c}\big| \,.
 \label{SVEA}
\end{align}
With this approximation, Eqs.~(\ref{nonlocal}) and (\ref{sigma12}) lead to
\begin{subequations}
\label{eqbasis1}
 \begin{align}
 &[\Delta_{12}(\textbf{v}) +i\gamma_{c}]\hat{\sigma}_{12}  + \Omega_{p}(\textbf{r})n_{0}F(\textbf{v})
 \nonumber \\
 & \qquad + \Omega_{c}\hat{\sigma}_{13}  -i \gamma_{c}\hat{R}_{12}(\textbf{r})F(\textbf{v})=0 \,,\label{eqbasis1a}\\[1ex]
  &\big[\Delta_{13}(\textbf{v})+i\gamma_{c}\big]\hat{\sigma}_{13} + \Omega_{c}\hat{\sigma}_{12}
   -i\gamma_{c}\hat{R}_{13}(\textbf{r},t)F(\textbf{v}) \nonumber \\
   &\qquad +n_{0}^{2}\Omega_{c}^{3}AF(\textbf{v})B(\textbf{r},\textbf{v})\Omega_{p}(\textbf{r})=0 \,, \label{eqbasis1b}
 \end{align}
\end{subequations}
where we have defined a kernel function depending on $\textbf{r}$ and $\textbf{v}$ as
\begin{align}
 B(\textbf{r},\textbf{v})=\int\frac{V(\textbf{r}-\textbf{r}')\big|\Omega_{p}(\textbf{r}')\big|^{2}d^{3}\textbf{r}'}{\Omega_{c}^{2}-\Delta_{12}(\textbf{v})\Delta_{13}(\textbf{v})+V(\textbf{r}-\textbf{r}')\Delta_{12}(\textbf{v})}\,.
\end{align}
%
Integrating Eq.~(\ref{eqbasis1}) over $\textbf{v}$, we find
\begin{subequations}
\label{eqbasis2}
 \begin{align}
 \textbf{k}_{p}\cdot\hat{\textbf{J}}_{12}(\textbf{r})=&(\Delta_{p}+i\gamma_{12})\hat{R}_{12}(\textbf{r})
  + n_{0}\Omega_{p}(\textbf{r})+\Omega_{c}\hat{R}_{13}(\textbf{r})\,,\\[2mm]
 \Delta\textbf{k}\cdot\hat{\textbf{J}}_{13}(\textbf{r})=&i\gamma_{13}\hat{R}_{13}(\textbf{r}) + \Omega_{c}\hat{R}_{12}(\textbf{r})
 +n_{0}^{2}\Omega_{c}^{3}A\Omega_{p}(\textbf{r})f_{1}(\textbf{r}) \,,
\end{align}
\end{subequations}
where we have set $\Delta=0$ for simplicity, and $\hat{\textbf{J}}_{\alpha\beta}(\textbf{r})$ and $f_{1}(\textbf{r})$ are defined as
\begin{align}
 \hat{\textbf{J}}_{\alpha\beta}(\textbf{r}) &= \int \hat{\sigma}_{\alpha\beta}(\textbf{r},\textbf{v})\textbf{v} d^{3}\textbf{v}\,, \\[1ex]
 f_{1}(\textbf{r}) &= \int F(\textbf{v})B(\textbf{r},\textbf{v})d^{3}\textbf{v}\,.
 \end{align}
In Eq.~(\ref{eqbasis1b}, when the residual Doppler effect $\Delta\text{k}\text{v}_{p}$ becomes dominant, we can expand $\hat{\sigma}_{13}(\textbf{r},\textbf{v})$ to the first order of $\Delta\text{k}\text{v}_{p}$ to give
\begin{align}
 \hat{\sigma}_{13}(\textbf{r},\textbf{v})\simeq\hat{R}_{13}(\textbf{r})F(\textbf{v}) +\frac{\hat{\sigma}^{(1)}_{13}(\textbf{r},\textbf{v})}{\Delta\text{k}\text{v}_{p}}\,,
 \label{expand}
\end{align}
which leads to
\begin{align}
 \hat{\textbf{J}}_{13} = \frac{1}{\Delta\text{k}\text{v}_{p}}\int\hat{\sigma}^{(1)}_{13}(\textbf{r},\textbf{v})\:d^{3}\textbf{v}\,.
\end{align}
Expanding $\hat{\sigma}_{13}(\textbf{r},\textbf{v})$ as in Eq.~(\ref{expand}), multiplying by $\textbf{v}$ in Eq(\ref{eqbasis1b}, integrating over $\textbf{v}$, and keeping the leading order in $\Delta\text{k}\text{v}_{p}$, we find 
\begin{align}
 &\Delta\text{k}^{2}\text{v}^{2}_{p}\hat{R}_{13}(\textbf{r})-i(\gamma_{12}+\gamma_{c})\Delta\textbf{k}\cdot\hat{\textbf{J}}_{13}(\textbf{r})\nonumber\\&=\Omega_{c}\Delta\textbf{k}\cdot\hat{\textbf{J}}_{12}(\textbf{r})+n_{0}^{2}\Omega_{c}^{3}A\Omega_{p}(\textbf{r})f_{2}(\textbf{r})\,.
\label{j13}
\end{align}
Here, $f_{2}(\textbf{r})$ is given by
\begin{align}
f_{2}(\textbf{r}) = \int F(\textbf{v})B(\textbf{r},\textbf{v})\Delta\textbf{k}\cdot\textbf{v}\:d^{3}\textbf{v}\,.
\end{align}
In experimental settings, a common choice of the laser field geometry is near-collinear propagation, i.e., $\textbf{k}_{p}\parallel\textbf{k}_{c}\parallel\Delta\textbf{k}$. Then from Eq.~(\ref{eqbasis2}) and (\ref{j13}), $\hat{R}_{13}(\textbf{r})$ can be obtained as
\begin{align}
& \hat{R}_{13}(\textbf{r})=\frac{\frac{\Delta\text{k}}{\text{k}_{p}}\Omega_{c}(\Delta_{p}+i\gamma_{12})+i\Omega_{c}(\gamma_{13}+\gamma_{c})}{\Delta\text{k}^{2}\text{v}_{p}^2-\frac{\Delta\text{k}}{\text{k}_{p}}\Omega_{c}^{2}+\gamma_{13}(\gamma_{13}+\gamma_{c})}\hat{R}_{12}(\textbf{r})\nonumber\\
& +\frac{\frac{\Delta\text{k}}{\text{k}_{p}}n_{0}\Omega_{c}\Omega_{p}(\textbf{r})+n_{0}^{2}\Omega_{c}^{3}A\Omega_{p}(\textbf{r})[f_{2}(\textbf{r})+i(\gamma_{c}+\gamma_{13})f_{1}(\textbf{r})]}{\Delta\text{k}^{2}\text{v}_{p}^2-\frac{\Delta\text{k}}{\text{k}_{p}}\Omega_{c}^{2}+\gamma_{13}(\gamma_{13}+\gamma_{c})}
 \label{R131}
\end{align}
In Eq.~(\ref{eqbasis1a}), we approximate $\hat{\sigma}_{13}(\textbf{r},\textbf{v})\simeq\hat{R}_{13}(\textbf{r})F(\textbf{v})$ and integrate over $\textbf{v}$, leading to
\begin{align}
\hat{R}_{13}(\textbf{r})=\frac{1}{\Omega_{c}}\bigg[-n_{0}\Omega_{p}(\textbf{r})+\frac{i\gamma_{c}G-1}{G}\hat{R}_{12}(\textbf{r})\bigg]\,,
\label{R132}
\end{align}
with
\begin{align}
 G = \int\frac{F(\textbf{v})}{\Delta_{p}-\textbf{k}_{p}\cdot\textbf{v}+i(\gamma_{12}+\gamma_{c})}\,.
\end{align}
Finally, the solution for $\hat{R}_{12}(\textbf{r})$ can be derived from Eq.~(\ref{R131}) and (\ref{R132}). The coherence $\rho_{21}(\textbf{r})$ which determine the atomic response reads

\begin{align}
 \rho_{21}(\textbf{r}) & =\langle\hat{R}_{12}(\textbf{r})\rangle \nonumber\\
 & =\frac{M}{D}\Omega_{p}(\textbf{r}) \nonumber\\
 &\quad+\frac{1}{D}n_{0}^{2}\Omega_{c}^{4}A\Omega_{p}(\textbf{r})\big[i(\gamma_{13}+\gamma_{c})f_{1}(\textbf{r})+f_{2}(\textbf{r})\big]\nonumber \\
 &=\frac{M}{D}\Omega_{p}(\textbf{r})+i\frac{(\gamma_{13}+\gamma_{c})n_{0}^{2}\Omega_{c}^{4}A}{D}\Omega_{p}(\textbf{r}) \nonumber \\
 &\qquad \int K(\textbf{r}-\textbf{r}')\big|\Omega_{p}(\textbf{r}')\big|^{2}d\textbf{r}'\,.
 \label{thermal}
\end{align}
with
\begin{subequations}
\begin{align}
 &D=\big[\Delta\text{k}^{2}\text{v}_{p}^{2}-\frac{\Delta\text{k}}{\text{k}_{p}}\Omega_{c}^{2}+\gamma_{13}(\gamma_{13}+\gamma_{c})\big]\frac{i\gamma_{c}G-1}{G}\nonumber \\
 &\qquad -\big[\frac{\Delta\text{k}}{\text{k}_{p}}(\Delta_{p}+\gamma_{12})+i(\gamma_{13}+\gamma_{c})\big]\Omega_{c}^{2} \,,\\[2mm]
&M=\big[\Delta\text{k}^{2}\text{v}_{p}^{2}+\gamma_{13}(\gamma_{13}+\gamma_{c})\big]n_{0} \,,\\[2mm]
& K(\textbf{r}) = \int\frac{\big(1-i\frac{\Delta\textbf{k}\cdot\textbf{v}}{\gamma_{13}+\gamma_{c}}\big)F(\textbf{v})V(\textbf{r})d^{3}\textbf{v}}{\Omega_{c}^{2}-\Delta_{12}(\textbf{v})\Delta_{13}(\textbf{v})+V(\textbf{r})\Delta_{12}(\textbf{v})}\,.
\end{align}
\end{subequations}
In the limit $T\rightarrow 0$, $\rho_{21}(\textbf{r})$ is reduced to
\begin{align}
 \rho_{21}(\textbf{r})&=\frac{i\gamma_{13}n_{0}}{\Omega_{c}^{2}-i\gamma_{13}\Delta_{12}}\Omega_{p}(\textbf{r})\nonumber \\
 & -\frac{n_{0}^{2}\Omega_{c}^{4}\Omega_{p}(\textbf{r})}{(\Omega_{c}^{2}-i\gamma_{13}\Delta_{12})|\Omega_{c}^{2}-i\gamma_{13}\Delta_{12}|^2}\nonumber \\
 &\times \int\frac{V(\textbf{r}-\textbf{r}')|\Omega_{p}(\textbf{r}')|^2}{\Omega_{c}^{2}-i\gamma_{13}\Delta_{12}+\Delta_{12}V(\textbf{r}-\textbf{r}')}d^{3}\textbf{r}'\,,
 \label{cold}
\end{align}
recovering the results for ultracold gases~\cite{Sevincli2011PRL}. Here, $\Delta_{12}=\Delta_{p}+i\gamma_{12}$.

In the paraxial regime, the propagation equation for the probe field finally becomes
\begin{align}
\left(\frac{\partial}{\partial z}-\frac{i}{2\text{k}_{p}}\frac{\partial^2}{\partial\textbf{r}_{\bot}^2}\right)\Omega_{p}(\textbf{r}_{\bot},z)
&=i\frac{3\lambda_{p}^2\Gamma_{21}}{8\pi}\rho_{21}(\textbf{r})
\label{prop-m}
\end{align}
From Eq.~(\ref{thermal}) and (\ref{prop-m}), we can examine the spatial evolution of the probe field.

\section{\label{app-a0}Derivation of $a_{0}(\zeta)$}

From Eq.~(\ref{eqa0}) the analytical solution for $a_{0}(\zeta)$ can be found by first rewriting $a_{0}(\zeta)=f(\zeta)e^{ig(\zeta)}$, which decomposes Eq.~(\ref{eqa0}) into two parts
\begin{subequations}
 \label{eqa0decomposition}
 \begin{empheq}[left=\empheqlbrace]{align}
  \frac{df(\zeta)}{d\zeta}&=-c_0^Rf^{3}(\zeta),\\
  \frac{dg(\zeta)}{d\zeta}&=c_0^If^{2}(\zeta),
\end{empheq}
\end{subequations}
where we have set $C_{nl}K_{F}(0)=c_{0}=c_0^R + i c_0^I$ with $c_0^R = \text{Re}[c_0]$ and $c_0^I = \text{Im}[c_0]$. Eq.~\ref{eqa0decomposition} can be solved to give
\begin{align}
  a_{0}(\zeta)&= (1 + 2c_0^I\zeta)^{\frac{ic_{0}}{2c_0^I}} \nonumber\\
 & = \frac{1}{\sqrt{1+2 c_0^I\zeta}} e^{i \frac{c_0^R}{2c_0^I} \ln(1 + 2c_0^I\zeta)}\,.
\end{align}

\section{\label{app-a1}Derivation of $a_1(\zeta)$}
We start by rewriting $a_{1}(\zeta)$ as 
\begin{align}
 a_{1}(\zeta)=\epsilon_{0} b_{1}(\zeta)e^{-\frac{i\text{k}^{2}_{x}\zeta }{2}+ i\frac{c_{0}+c_{1}}{2c_0^I}\text{ln}(1+2c_0^I\zeta)},
\end{align}
such that Eq.~(\ref{a1simplified}) becomes
\begin{align}
 \frac{db_{1}(\zeta)}{d\zeta} = \frac{ic_{1}b^{*}_{1}(\zeta)}{1+2c_0^I\zeta}e^{i\text{k}^{2}_{x}\zeta-i\frac{c_1^R}{c_0^I}\text{ln}(1+2c_0^I\zeta)},
 \label{eqb1}
\end{align}
Taking the complex conjugate of Eq.~(\ref{eqb1}) and to replace $b^{*}_{1}$ leads to
\begin{align}
 \frac{d^{2}b_{1}}{d\zeta^2} = &\bigg[i\text{k}^{2}_{x} - \frac{2(c_0^I+ic_1^R)}{1+2c_0^I\zeta}\bigg]\frac{db_{1}}{d\zeta}\nonumber \\
 & + \frac{|c_{1}|^2b_{1}}{(1+2c_0^I\zeta)^2}.
 \label{eqb1final}
\end{align}
This way, the solution of $b_{1}(\zeta)$ can be obtained  as
\begin{align}
 b_{1}(\zeta) &= \big(1+2c_0^I\zeta\big)^{\frac{-ic_{1}}{2c_0^I}} \left[ s_{1}U\left( \frac{-ic_{1}}{2c_0^I}, \right. \right.
 \nonumber \\
&  \left. 1+\frac{c_1^I}{c_0^I},\frac{i\text{k}^{2}_{x}}{2c_0^I}+i\text{k}^{2}_{x}\zeta\right)  \nonumber \\
 & \left. +s_{2}L \left(\frac{ic_{1}}{2c_0^I},\frac{c_1^I}{c_0^I},\frac{i\text{k}^{2}_{x}}{2c_0^I}+i\text{k}^{2}_{x}\zeta \right ) \right ] \,,
\end{align}
with the confluent hypergeometric function $U(a,b,z)$ and Laguerre polynomials $L(n,a,x) \equiv L^{a}_n(x)$. The coefficients $s_{1}$ and $s_{2}$ are constrained by the initial conditions
\begin{subequations}
\begin{align}
  b_{1}(\zeta=0)&=1,\\ 
  \frac{db_{1}}{d\zeta}\bigg\rvert_{\zeta=0} &= ic_{1}\,.
  \end{align}
  \label{initial-conditions}
\end{subequations}
Finally we obtain
\begin{align}
 a_{1}(\zeta) &= \epsilon_{0}\big(1+2c_0^I\zeta\big)^{\frac{ic_{0}}{2c_0^I}}e^{-\frac{i\text{k}^{2}_{x}\zeta}{2}}  \left[ s_{1}U\left( \frac{-ic_{1}}{2c_0^I}, \right. \right.
 \nonumber \\
&  \left. 1+\frac{c_1^I}{c_0^I},\frac{i\text{k}^{2}_{x}}{2c_0^I}+i\text{k}^{2}_{x}\zeta\right)  \nonumber \\
 & \left. +s_{2}L \left(\frac{ic_{1}}{2c_0^I},\frac{c_1^I}{c_0^I},\frac{i\text{k}^{2}_{x}}{2c_0^I}+i\text{k}^{2}_{x}\zeta \right ) \right ] \,.
\end{align}
\section{\label{app-a1-case1}Derivation of $a_1(\zeta)$ when $\text{Im}[C_{nl}K_{F}(0)]=0$}
In the case of $\text{Im}[C_{nl}K_{F}(0)]$, we find from Eq.~(\ref{eq-decomposition}) that 
\begin{align}
 a_{0}(\zeta) = e^{i[-\frac{\text{k}^{2}_{0}}{2} + c^{R}_{0}]\zeta}\,.
\end{align}
Thus the equation of motion for $a_{1}$ and $a_{-1}$ is reduced to 
\begin{subequations}
\label{eq-case1-general}
 \begin{align}
  \frac{da_{1}(\zeta)}{d\zeta} = \;&i\big[-\frac{\text{k}^{2}_{x}}{2}+c^{R}_{0}+c_{1}\big]a_{1}(\zeta) + ic_{1}e^{i[-\text{k}^{2}_{0} + 2c^{R}_{0}]\zeta}a^{*}_{-1}(\zeta)\,,\\
  \frac{da_{-1}(\zeta)}{d\zeta} = \;&i\big[-\frac{(2\text{k}_{0}-\text{k}_{x})^2}{2}+c^{R}_{0}+c_{-1}\big]a_{-1}(\zeta) \nonumber\\
  &+ ic_{-1}e^{i[-\text{k}^{2}_{0} + 2c^{R}_{0}]\zeta}a^{*}_{1}(\zeta)\,.
 \end{align}
\end{subequations}
where we have set $c_{0}=C_{nl}K_{F}(0),c_{1}=C_{nl}K_{F}(\text{k}_{x}-\text{k}_{0})$ and $c_{-1}=C_{nl}K_{F}(\text{k}_{0}-\text{k}_{x})$. With the transformation
\begin{align}
a_{1}(\zeta) = \;& b_{1}(\zeta)e^{i\big[-\frac{\text{k}^{2}_{x}}{2}+c^{R}_{0}+c_{1}\big]\zeta}\,,\\
a_{-1}(\zeta) = \;& b_{-1}(\zeta)e^{i\big[-\frac{(2\text{k}_{0}-\text{k}_{x})^2}{2}+c^{R}_{0}+c_{-1}\big]\zeta}\,,
\end{align}
Eq.~(\ref{eq-case1-general}) can be simplified to 
\begin{align}
 \frac{db_{1}(\zeta)}{d\zeta}=ic_{1}b^{*}_{-1}(\zeta)e^{i\big[(\text{k}_{x}-\text{k}_{0})^2-c_{1}-c^{*}_{-1}\big]\zeta}\,,\label{b1-case1-general}\\
 \frac{db_{-1}(\zeta)}{d\zeta}=ic_{-1}b^{*}_{1}(\zeta)e^{i\big[(\text{k}_{x}-\text{k}_{0})^2-c_{-1}-c^{*}_{1}\big]\zeta}\,,\label{bm1-case1-general}
\end{align}

Replacing $b_{-1}(\zeta)$ in Eq.~(\ref{bm1-case1-general}) by $b_{-1}(\zeta)$ given in Eq.~(\ref{b1-case1-general}) leads to 
\begin{align}
 \frac{d^2 b_{1}(\zeta)}{d\zeta^{2}} = \;& i\bigg[(\text{k}_{x}-\text{k}_{0})^2 -c_{1}-c^{*}_{-1}\bigg]\frac{d b_{1}(\zeta)}{d\zeta}\nonumber\\
 &+ c_{1}c^{*}_{-1}b_{1}(\zeta)\,.
 \label{b1-case1-final}
\end{align}
From Eq.~(\ref{b1-case1-final}) $b_{1}(\zeta)$ can be easily obtained. Finally, the exact expression for $a_{1}(\zeta)$ can be written as
\begin{align}
 a_{1}(\zeta) =\;& \epsilon_0 e^{i\big[\frac{\text{k}_{0}(\text{k}_{0}-2\text{k}_{x})}{2}+ c^{R}_{0}+\frac{c_{1}-c^{*}_{-1}}{2}\big]\zeta}\nonumber\\
                 &\times\bigg[s^{'}_{1}e^{\lambda^{'}(\text{k}_{x})\zeta} + s^{'}_{2}e^{-\lambda^{'}(\text{k}_{x})\zeta} \bigg]\,,
\end{align}
with 
\begin{align}
 \lambda^{'}(\text{k}_{x}) = \frac{1}{2}\sqrt{4c_{1}c^{*}_{-1} - \big[(\text{k}_{x}-\text{k}_{0})^2-c_{1}-c^{*}_{-1}\big]^2}\,.
\end{align}
The coefficient $s^{'}_{1}$ and $s^{'}_{2}$ are defined by the initial conditions. With the same procedures, $a_{-1}(\zeta)$ can also be attained, which is not shown here. For the specific case when $\text{k}_{0} = 0$, we have $c_{1}=c_{-1}$ as we have discussed in the main text. Thus $a_{1}(\zeta)$ is reduced to 
\begin{align}
 a_{1}(\zeta) =\;& \epsilon_0 e^{ic^{R}_{0}\zeta-c^{I}_{1}\zeta}
                 \bigg[s^{'}_{1}e^{\lambda^{'}(\text{k}_{x})\zeta} + s^{'}_{2}e^{-\lambda^{'}(\text{k}_{x})\zeta} \bigg]\,,
\end{align}
and
\begin{align}
 \lambda^{'}(\text{k}_{x}) = \frac{1}{2}\sqrt{4|c_{1}|^2-(\text{k}^{2}_{x}-2c^{R}_{1})^2}\,.
\end{align}

\bibliographystyle{apsrev4-1}
\bibliography{referencesbase}

\end{document}